\documentclass[jou]{interact}

\usepackage{lineno}
\usepackage{dsfont}
\usepackage{xcolor}
\usepackage{tcolorbox}
\usepackage{cancel}

\usepackage{array}
\usepackage{booktabs}
\usepackage[english]{babel} 
\usepackage{inputenc}

\usepackage{multicol, amsthm}

\usepackage{parskip}
\usepackage[noframe]{showframe}
\usepackage{framed}
\usepackage{lipsum}
\renewenvironment{shaded}{%
  \MakeFramed{\advance\hsize-\width \FrameRestore\FrameRestore}}%
 {\endMakeFramed}
\definecolor{shadecolor}{gray}{0.75}

\usepackage{booktabs,array}

\newcount\rowc

\makeatletter
\def\ttabular{%
\hbox\bgroup
\let\\\cr
\def\rulea{\ifnum\rowc=\@ne \hrule height 1.3pt \fi}
\def\ruleb{
\ifnum\rowc=1\hrule height 1.3pt \else
\ifnum\rowc=6\hrule height \heavyrulewidth 
   \else \hrule height \lightrulewidth\fi\fi}
\valign\bgroup
\global\rowc\@ne
\rulea
\hbox to 10em{\strut \hfill##\hfill}%
\ruleb
&&%
\global\advance\rowc\@ne
\hbox to 10em{\strut\hfill##\hfill}%
\ruleb
\cr}
\def\endttabular{%
\crcr\egroup\egroup}

\usepackage{epstopdf}
\usepackage[caption=false]{subfig}

\usepackage[doublespacing]{setspace}
\usepackage{inputenc}

\theoremstyle{plain}

\theoremstyle{definition}

\theoremstyle{remark}

\usepackage{natbib}
\begin{document}


\title{The consequences of checking for zero-inflation and overdispersion in the analysis of count data}

\author{
\name{Harlan Campbell, harlan.campbell@stat.ubc.ca}
}

\maketitle
\vspace{-0.25cm}
\begin{abstract}
\noindent 1.  Count data are ubiquitous in ecology and the Poisson generalized linear model (GLM) is commonly used to model the association between counts and explanatory variables of interest.  \textcolor{black}{When fitting this model to the data, one typically proceeds by first confirming that the model assumptions are satisfied.  If the residuals appear to be overdispersed or if there is zero-inflation, key assumptions of the Poison GLM may be violated and }researchers will then typically consider alternatives to the Poison GLM.  An important question is whether the potential model selection bias introduced by this data-driven multi-stage procedure merits concern.

\noindent 2.  Here we conduct a large-scale simulation study to investigate the potential consequences of model selection bias that can arise in the simple scenario of analyzing a sample of potentially overdispersed, potentially zero-heavy, count data.  \textcolor{black}{Specifically, we investigate model selection procedures recently recommended by \cite{blasco2019does}} using either a series of score tests or the AIC statistic to select the best model.

\noindent 3.  We find that, when sample sizes are small, model selection based on preliminary score tests (or the AIC) can lead to potentially substantial type 1 error inflation.  When sample sizes are large,  model selection based on preliminary score tests is less problematic.

\noindent 4.  Ignoring the possibility of overdispersion and zero inflation during data analyses can lead to invalid inference.  However, if one does not have sufficient power to test for overdispersion and zero inflation, \emph{post hoc} model selection may also lead to substantial bias.  This ``catch-22'' suggests that, if sample sizes are small,  a healthy skepticism is warranted whenever one rejects the null hypothesis.

\end{abstract}

\newpage

\section{Introduction}

\vskip 0.22in

Despite the ongoing debate surrounding the use (and misuse) of significance testing in ecology \citep{murtaugh2014defense, dushoff2019can} (and in other fields \citep{amrhein2019scientists}), hypothesis testing remains prevalent.  Indeed, many research fields have been criticized for publishing studies with serious errors of testing and interpretation, and ecologists have been accused of being ``confused'' about  when and how to conduct appropriate hypothesis tests \citep{stephens2005information}.  One issue that receives a substantial amount of attention is that of failing to check for possible violations of distributional assumptions.  According to \citet{freckleton2009seven},  using statistical tests that assume a given distribution on the data while failing to test for the assumptions required of said  distribution is one of ``seven deadly sins.''

One of the most popular statistical models in ecology (and in many other fields, e.g., finance, psychology, neuroscience, and microbiome research, \citep{bening2012generalized, loeys2012analysis, zoltowski2018scaling, xu2015assessment}) is the Poisson generalized linear model (GLM) \citep{nelder1972generalized}.  With count outcome data, a Poisson GLM is the most common starting point for testing an association between a given outcome, $Y$, and a given covariate of interest, $X$. The Poisson GLM assumes the outcome data, conditional on the covariates, are the result of independent sampling from a Poisson distribution where, importantly, the mean and variance are equal.  However, in practice, count data will often be show more variation than is implied by the Poisson distribution and the use of Poisson models is not always appropriate \citep{cox1983some}.


Count data frequently exhibit two (related) characteristics: (1) overdispersion and (2) zero-inflation. \textcolor{black}{Overdispersion refers to an excess of variability, while zero-inflation refers to an excess of zeros \citep{yang2010score}.  If model residuals} are overdispersed or have an excess of zeros, assumptions underlying the Poisson GLM will not hold and ignoring this will lead to serious errors (e.g., biased parameter estimates and invalid standard errors). It is therefore routine practice for researchers to check if the assumptions required of \textcolor{black}{a} Poisson model hold and adopt an alternative statistical model in the event that they do not; see \citet{zuur2010protocol}. 

 In the case of overdispersion, popular alternatives to the Poison GLM include the Quasi-Poisson (QP) model \citep{wedderburn1974quasi} and the Negative Binomial (NB) model  \citep{richards2008dealing, linden2011using}. Note that when selecting between the QP and NB models, the best choice is not always straightforward; see \citet{ver2007quasi}.  In the case of zero-inflation, popular alternatives to the Poison GLM include the Zero-Inflated Poisson model (ZIP) \citep{martin2005zero, lambert1992zero} and the Zero-Inflated Negative-Binomial model (ZINB) \citep{greene1994accounting}. 


  A multi-stage procedure will typically have researchers testing for overdispersion and zero-inflation in a preliminary stage (based on the residuals from a Poisson GLM), before testing the main hypothesis of interest (i.e., the association between $Y$ and $X$) in a second stage; see \citet{blasco2019does}.  If the first stage tests are not significant, the Poisson GLM is selected, regression coefficients are estimated along with their standard errors, and $p$-values are calculated allowing one to test for the association between $Y$ and $X$.  On the other hand, if the first stage test for overdispersion is significant, a QP or a NB model will be fit to the data.  Or, alternatively, if the first stage test for zero-inflation is significant, a ZIP model may be used.  In cases when there is evidence of both overdispersion and zero-inflation, more complex models such as the ZINB model or hurdle models will often be considered; see \citet{zorn1998analytic}. 
  
  Such a multi-stage, multi-test procedure may appear rather reasonable, and goodness-of-fit tests are frequently reported to confirm that the model-selection is appropriate.  However, recently, some researchers have warned against preliminary testing for distributional assumptions; e.g., \citet{shuster2005diagnostics} and \cite{wells2007dealing}. Their warnings are based on the following concern: since the preliminary tests are applied to the same data as the main hypothesis tests, this multi-stage procedure amounts to ``using the data twice'' \citep{hayes2020}. A hypothesis test using a model selected based on preliminary testing fails to take into account one's uncertainty with regards to the distributional properties of the data. \textcolor{black}{Unless the preliminary tests and the main hypothesis tests are entirely independent, this can result in model selection bias.} 
  
  The model selection bias at issue here is not the better known model selection bias associated with deciding \emph{post hoc} which variables to include in the model, e.g., the model selection bias associated with stepwise regression \citep{hurvich1990impact, whittingham2005habitat}.  Instead, here we are concerned with the potential bias introduced when deciding \emph{post hoc} which distributional assumptions should be accepted. The implications of considering \emph{post hoc} alternatives (or adjustments) to accommodate for distributional assumptions have been previously considered in other contexts.  Three examples come to mind.
    
    
      First, in the context of time-to-event data, the consequences of checking and adjusting for potential violations of the proportional hazards (PH) assumption required of a Cox PH model are considered by \citet{campbell2014consequences}.  The authors find that the ``common two-stage approach'' (in which one selects a model based on a preliminary test for PH) can lead to a substantial inflation of the type 1 error, even in scenarios where there is no violation of the PH assumption.  
     
     Second, in the simple context of testing the means of two independent samples, \citet{rochon2012test} investigate the consequences of conducting a preliminary test for normality (e.g., the Shapiro-Wilk test).  The authors conclude that while ``[f]rom a formal perspective, preliminary testing for normality is incorrect and should therefore be avoided,'' in practice, ``preliminary testing does not seem to cause much harm.''  
     
     
     Finally, in the context of clinical trials, factorial trials are an efficient method of estimating multiple treatments in a single trial.   However, factorial trials rely on the strict assumption of no interaction between the different treatments. \citet{kahan2013bias} investigates the consequences of conducting a preliminary test for the interaction between treatment arms (as is often recommended).  By means of a simulation study, \citet{kahan2013bias} shows that the estimated treatment effect from a factorial trial under the ``two-stage analysis'' can be severely biased, even in the absence of a true interaction.
     
  Model selection bias is considered a ``quiet scandal in the statistical community'' \citep{breiman1992little} and is now all the more important to understand given recent concerns with research reproducibility and researcher incentives \citep{kelly2019, nosek2012scientific, gelman2013garden, fraser2018questionable}.  In some fields, such as psychology, the issue is finally being recognized.  \citet{williams2019dealing} conclude that ``it is currently unclear how [psychology] researchers should deal with distributional assumptions'' since ``diagnosing and responding to distributional assumption problems'' may result in ``error rates [that] vary considerably from the nominal error rates.''  
  
  In ecology, some have warned about model selection bias (e.g., \citet{buckland1997model}), but the problem ``remains widely over-looked'' \citep{whittingham2006we}.  Indeed, ecologists will readily admit that ``this problem is commonly not appreciated in modelling applications'' \citep{whittingham2005habitat}.  \citet{anderson2007model} notes that: ``Model selection bias is subtle but its effects are widespread and little understood by many people working in the life sciences.''  
  
  {In this paper, we conduct a large-scale simulation study to investigate the potential consequences of model selection bias that can arise in the simple scenario of analyzing a sample of potentially overdispersed, potentially zero-inflated, count data.} \textcolor{black}{It is difficult to anticipate what these consequences might be.  Often, while model selection bias is problematic from a theoretical perspective, it does not lead to substantial problems in practice.  We restrict our attention to two model selection procedures, one based on conducting score tests, and another based on calculating AIC statistics.}  
  
  In Section 2, we review commonly used models and in Section 3, we outline the framework of a simulation study to investigate the consequences of checking for zero-inflation and overdispersion.  In Section 4, we discuss the results of this simulation study and we conclude in Section 5 with a summary of findings and general recommendations.

\section{Materials and Methods}

\subsection{Models for the analysis of count data}

Let us begin with the simplest version of the Poisson GLM.  Let $Y_{i}$, for $i$ in $1, \ldots, n$, be the outcome of interest observed for $n$ independent samples. Let $X_{i}$, for $i$ in $1,\ldots,n$, represent a single covariate of interest.  If the covariate of interest is categorical with $k$ different categories (e.g., $k$ different species of fish), $X_{i}$ will be a vector with length equal to $k-1$;  otherwise it will be a single scalar and $k=2$.  The simplest Poisson regression model, with a standard $log$ link, will have that:
\begin{equation}
    Y_{i} \sim Poisson(\lambda_{i} = exp(\beta_{0} + \beta_{X}X_{i})), \quad \textrm{or equivalently:}
\end{equation}
\textcolor{black}{
\begin{equation}
    Pr(Y_{i}=y_{i}|\beta_{0}, \beta_{X}) = \frac{(exp(\beta_{0} + \beta_{X}X_{i}))^{y_{i}}exp(-exp(\beta_{0} + \beta_{X}X_{i}))}{y_{i}!}, 
\end{equation}
}
\noindent for   $i  \textrm{ in }1,\ldots,n$, where $\beta_{0}$ is the intercept, and $\beta_{X}$ is the coefficient (or coefficient-vector of length $k-1$) representing the association between $X$ and $Y$. Note that this model implies the following equality: $E(Y_{i}) = Var(Y_{i}) = \lambda_{i}$, for  $i$ in $1,\ldots, n$.  

Parameter estimates, $\widehat{\beta_{0}}$, and $\widehat{\beta_{X}}$, can be obtained by maximum likelihood estimation via iterative Fischer scoring.  A confidence interval for $\beta_{X}$ is typically calculated by the standard profile likelihood approach where one inverts a likelihood-ratio test; see \citet{venzon1988method}, or more recently \cite{uusipaikka2008confidence}.  Maximum likelihood estimation via iterative Fischer scoring is implemented as a default for the \verb|glm()| function in \verb|R| \citep{dunn2018generalized} and profile likelihood confidence intervals are provided by default when using the \verb|confint()| function for GLMs; see \citet{ripley2013package}.

To test whether there is an association between $Y$ and $X$, we define the following hypothesis test: $H_{0}: \beta_{X} = 0$ vs. $H_{1}: \beta_{X} \ne 0$.  A simple likelihood ratio test (LRT), or Wald test will provide a $p$-value to evaluate this hypothesis; see \citet{zeileis2008regression}.  The LRT and Wald test are asymptotically equivalent.   For the likelihood ratio test, the $Z$ statistic is obtained by calculating the null and residual deviance as $Z_{LRT} = D_{1}-D_{0}$, where:

\(D_{0} =2 \sum_{i=1}^{n}\left\{Y_{i} \log \left(Y_{i} / exp(\widehat{\beta}_{0}) \right)-\left(Y_{i}-exp(\widehat{\beta}_{0})\right)\right\}\), \textrm{and:} \\

\textcolor{black}{
\(D_{1} =2 \sum_{i=1}^{n}\left\{Y_{i} \log \left(Y_{i} / \widehat{{\lambda}_{i}}\right)-\left(Y_{i}-\widehat{{\lambda}_{i}}\right)\right\}\),  where $\widehat{\lambda_{i}}=\exp (\widehat{\beta}_{0}+\widehat{\beta}_{X} X_{i})$.
}


 If the distributional assumptions of the Poisson GLM are met and the null hypothesis holds, the $Z$ statistic will follow (asymptotically) a $\chi^{2}$ distribution with $df = k-1$ degrees of freedom, and the $p$-value is calculated as: $ p\textrm{-value} = P_{\chi^{2}}(Z, df = k-1)$. (For the Wald test, with $k=2$, the $Z$-statistic is defined as $Z_{Wald} = \left({\widehat{\beta_{X}}}/{\textrm{se}(\widehat{\beta_{X}})}\right)^{2}$, where $\textrm{se}(\widehat{\beta_{X}})$ is the standard error of the maximum likelihood estimate (MLE); see \cite{hilbe20077} for details when $k>2$).  


 However, if the distributional assumptions do not hold, the $Z$ statistic will be compared with the wrong reference distribution invalidating any significance test (and associated confidence intervals).   {Therefore, in order to conduct valid inference, researchers will typically carry out an extensive model selection procedure.  \textcolor{black}{Note that model selection must always be based on model residuals and not on the distribution of the response variable (which is erroneously done on occasion).} \citet{blasco2019does} outline and illustrate a proposed protocol.  Such a procedure is typically based on:

\begin{itemize}
    \item measuring indices (e.g., the dispersion index \citep{fisher1950significance}; the zero-inflation index \citep{puig2006count});
    
    \item conducting score tests (e.g., the $D\&L$ score test for Poisson vs. NB regression \citep{dean1989tests}; the $vdB$ score test for Poisson vs. ZIP regression \citep{van1995score}; the score test for ZIP vs ZINB regression \citep{ridout2001score});
    
    \item and evaluating candidate models with goodness-of-fit tests (e.g., likelihood ratio tests; the Vuong and Clarke tests) and model selection criteria (e.g., AIC, AICc and BIC).
\end{itemize}
}
   
    In this paper, for simplicity, we will only consider three alternative models \textcolor{black}{in addition to the Poisson model described above}: the (type 2) NB, the ZIP, and the (type 2) ZINB regression models as described in \citet{blasco2019does}.  Let us briefly review these three alternative regression models.

       \noindent \textbf{(1) The ZIP regression model - }  We will consider the  following zero-inflated Poisson model where the probability of a structural zero, $\omega_{i}$, is a function of the covariate $X_{i}$.  Specifically,     
       \begin{eqnarray}
           Pr(Y_{i}=y_{i}|\omega_{i}, \lambda_{i}) &=& \omega_{i} + (1-\omega_{i})exp(-\lambda_{i}), \quad \textrm{if} \quad y=0; \\
           &=& (1-\omega_{i})exp(-\lambda_{i})\lambda_{i}^{y_{i}}/y_{i}!, \quad \textrm{if} \quad  y_{i} > 0;  \nonumber 
       \end{eqnarray}
       \noindent where we have a log link function for $\lambda_{i}$ and a logit link function for $\omega_{i}$ (for $i$ in $1,\ldots,n$) such that: 
      \begin{eqnarray}
      \label{eq:log_logit1}
       \lambda_{i} &=& exp(\beta_{0} + \beta_{X}X_{i}), \quad \textrm{and} \\
            \label{eq:log_logit2}
       \omega_{i} &=& \left(\frac{exp(\gamma_{0} + \gamma_{X}X_{i})}{1+exp(\gamma_{0} + \gamma_{X}X_{i})}\right).
        \end{eqnarray}      
       The ZIP model has that  $0 \le \omega_{i} \le 1$  and $\lambda_{i} > 0$, and implies the following about the mean and variance of the data: $\textrm{E}(Y_{i}) = \lambda_{i}(1-\omega_{i}) = \mu_{i}$ and $\textrm{Var}(Y_{i}) = \mu_{i} +  \mu_{i}^2\omega_{i}/(1-\omega_{i})$.  \textcolor{black}{ The dispersion index is therefore equal to $d = \textrm{Var}(Y_{i})/\textrm{E}(Y_{i}) = 1 + \lambda_{i}\omega_{i}$.   As $\omega_{i} \rightarrow 0$, we have that $Y_{i}$ reverts to follow a Poisson distribution with mean $\lambda_{i}$. } A null hypothesis of no association between $X$ and $Y$ is specified as: $H_{0} : \beta_{X} = \gamma_{X} = 0$.
       
       
  \noindent \textbf{(2) The (type 2) NB regression model - }  We will consider the following NB regression model:    
%
 \begin{eqnarray}
\Pr(Y_{i} = y_{i}|\nu,\lambda_{i}) = \frac{\Gamma(y_{i} + \nu)}{\Gamma(y_{i} + 1)\Gamma(\nu)}\left(\frac{1}{1 + \lambda_{i}/\nu}\right)^{\nu}\left(\frac{\lambda_{i}/\nu}{1 + \lambda_{i}/\nu} \right)^{y_{i}};
\end{eqnarray}
\noindent where we use a log link function for $\lambda_{i} = exp(\beta_{0} + \beta_{X}X_{i})$, and where $\nu > 0$ is a dispersion parameter that does not depend on covariates.  The type 2 NB model implies the following about the mean and variance of the data: $\textrm{E}(Y_{i}) = \lambda_{i}$, and  $\textrm{Var}(Y_{i}) = \lambda_{i} + \lambda_{i}^{2}/\nu$. \textcolor{black}{  The dispersion index is therefore equal to $d = Var(Y_{i})/E(Y_{i}) = 1 + \lambda_{i}/\nu$.}  As $\nu \rightarrow \infty$, we have that $Y_{i}$ reverts to follow a Poisson distribution with mean $\lambda_{i}$.  A null hypothesis of no association between $X$ and $Y$ is specified as: $H_{0} : \beta_{X} = 0$.

 \noindent \textbf{(3) The (type 2) ZINB regression model - }   We will consider the  following ZINB regression model:
       \begin{eqnarray}
           Pr(Y_{i}=y_{i}|\nu, \omega_{i}, \lambda_{i}) &=& \omega_{i} + (1-\omega_{i})(1/(1 + \lambda_{i}/\nu))^{\nu}, \quad \textrm{if} \quad y=0; \\
           &=& (1-\omega_{i})\frac{\Gamma(y_{i} + \nu)}{\Gamma(y_{i} + 1)\Gamma(\nu)}\left(\frac{1}{1 + \lambda_{i}/\nu}\right)^{\nu}\left(\frac{\lambda_{i}/\nu}{1 + \lambda_{i}/\nu} \right)^{y_{i}}, \quad \textrm{if} \quad  y_{i} > 0;  \nonumber 
       \end{eqnarray}
 \noindent where we use a log link function for $\lambda_{i}$ and a logit link function for $\omega_{i}$ as described in equations (\ref{eq:log_logit1}) and (\ref{eq:log_logit2}); and where $\nu > 0$ is a dispersion parameter that does not depend on covariates.  The type 2 ZINB model implies the following about the mean and variance of the data: $\textrm{E}(Y_{i}) = \lambda_{i}(1-\omega_{i})$, and  $\textrm{Var}(Y_{i}) =  (1-\omega_{i})(\lambda_{i} + \lambda_{i}^{2}(\omega_{i} + 1/\nu))$. \textcolor{black}{  The dispersion index is therefore equal to $d = Var(Y_{i})/E(Y_{i}) = 1 + \lambda_{i}(\omega_{i} + 1/\nu)$.}  A null hypothesis of no association between $X$ and $Y$ is specified as: $H_{0} : \beta_{X} = \gamma_{X} = 0$.

\subsection{Simulation Study}
\label{sec:methods}

As discussed in the previous section, prevailing practice for the analysis of count data is first to try to fit one's data with a Poisson GLM and only consider alternatives in the event that a preliminary test indicates that the distributional assumptions of the Poisson GLM may be violated. We will therefore consider the following multi-stage testing procedure in our simulation study investigation.  This follows the recommendations of \citet{blasco2019does} yet represents a simplification of the typical process followed by researchers.  \cite{walters2007using} also recommends a similar multi-step model selection procedure.  

For the illustrative purposes of this paper, we consider the \citet{dean1989tests} score test (D\&L test) for oversdispersion and the \cite{vuong1989likelihood} test for zero-inflation (see Appendix for details) in the following seven step procedure:

\pagebreak

\begin{shaded}
\begin{itemize}
    \item  \textbf{Step 1.}  Conduct the $D\&L$ score test for overdispersion ($H_{0}$: Poisson vs. $H_{1}$: NB).

    \item{ \begin{itemize}
        \item \textbf{Step 2.} If the $D\&L$ score test fails to reject the null, conduct a Vuong test for zero-inflation ($H_{0}$: Poisson vs. $H_{1}$: ZIP).  Otherwise, proceed to Step 5.
        
        \item{ \begin{itemize}
            \item \textbf{Step 3.} If the Vuong test for zero-inflation fails to reject the null, fit the Poisson GLM and calculate the $p$-value ($H_{0}: \beta_{X} = 0$ vs. $H_{1}: \beta_{X} \ne 0$).  Otherwise, proceed to Step 4.
            \item \textbf{Step 4.} If the Vuong test for zero-inflation rejects the null, fit the ZIP model and calculate the $p$-value ($H_{0} : \beta_{X} = \gamma_{X} = 0$).
        \end{itemize}}
        
         \item \textbf{Step 5.} If the $D\&L$ score test rejects the null, conduct the Vuong test for zero-inflation ($H_{0}$: NB vs. $H_{1}$: ZINB).
         
           \item{ \begin{itemize}
            \item \textbf{Step 6.} If the the Vuong test for zero-inflation fails to reject the null, fit the NB model and calculate the $p$-value ($H_{0} : \beta_{X} = 0$).  Otherwise, proceed to Step 7.
            \item \textbf{Step 7.} If the Vuong test for zero-inflation rejects the null, fit the ZINB regression model and calculate the $p$-value ($H_{0} : \beta_{X} = \gamma_{X} = 0$).
        \end{itemize}}
         
    \end{itemize}}
\end{itemize}

\end{shaded}

\begin{figure}
    \centering
    \includegraphics[width=12cm]{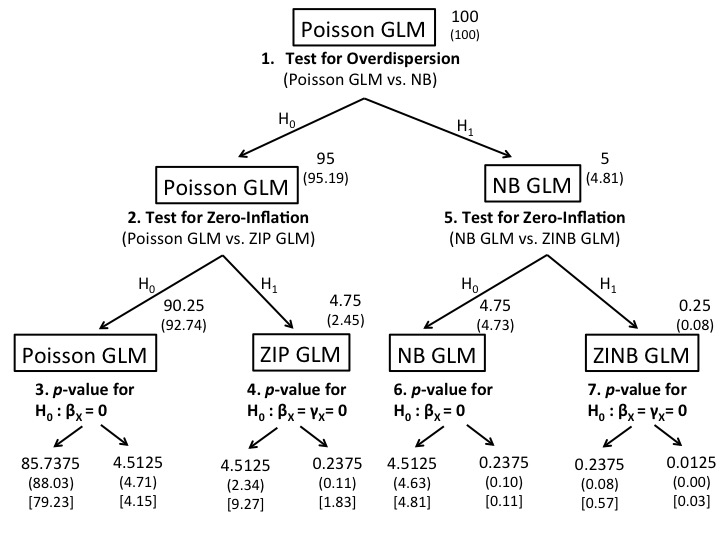}
    \caption{The multi-stage model selection procedure.  The Poisson GLM is the starting point.  Three score tests lead to one of four models.  Numbers in the top right-hand corner of each node indicate the expected number of datasets (out of a total of 100) to reach each outcome if the data was Poisson (with $\beta_{X}=0$), and each of the tests were truly independent (with a $\alpha=0.05$ type 1 error rate).   Numbers in parentheses correspond to results from the simulation study (Scenario ``3'', with $\phi=\infty$, $\omega=0$, $\beta_{0}=0.5$ and $n=250$).  Numbers in (curved) parentheses are those obtained in following the seven-step procedure; numbers in [square] parentheses are those obtained via AIC.  The unconditional type 1 error rate obtained in the simulation study following the seven-step procedure is 4.92\% (= 4.71 + 0.11 + 0.10 + 0.00).    The unconditional type 1 error rate obtained the simulation study when selecting the best model via AIC is 6.12\% (= 4.15 + 1.83 + 0.11 + 0.03).}
    \label{fig:tree}
\end{figure}

 Figure \ref{fig:tree} illustrates the multi-stage model selection procedure with the Poisson GLM as the starting point. Note that, in their example analysis of plant-herbivore interaction data,  \cite{blasco2019does} conduct a version of the above procedure.  First, based on the $D\&L$ score test, ``the data is clearly overdispersed and a NB model was preferred to a Poisson.''   The authors also conduct  Vuong and Clarke tests: ``The Vuong and Clarke tests rejected the Poisson and NB models in favour of their zero‐inflated versions[...].''  We decided to consider the Vuong test in our simulations instead of the Clarke test (or the Ridout score test), since the Vuong test appears to be the most widely used in practice.  \textcolor{black}{We also investigate another, simpler, model selection strategy: among the four models considered, the model with lowest $AIC$ is chosen and the corresponding $p$-value for the association between $X$ and $Y$ is calculated \citep{brooks2019statistical}.}

We conducted a large-scale simulation study in which samples of data were drawn from four different distributions:

\begin{enumerate}
    \item the Poisson distribution: \\
          \indent  $y_{i} \sim Poisson(\lambda = exp(\beta_{0}))$, for $i$ in 1,...$n$;
 
    \item the (type 2) Negative Binomial distribution: \\
    \indent  $y_{i} \sim NegBin(\nu, \lambda = exp(\beta_{0}))$, for $i=1,...,n$;
    
    \item the Zero-Inflated Poisson distribution: \\
    \indent $y_{i} \sim ZIPoisson(\omega, \lambda = exp(\beta_{0}))$, for $i=1,...,n$; and

        \item the Zero-Inflated Negative Binomial distribution: \\
        \indent $y_{i} \sim ZINegBin(\nu, \omega, \lambda = exp(\beta_{0}))$,
       for $i=1,...,n$.

\end{enumerate}

\noindent For each scenario, all data are simulated under the null hypothesis (i.e., with $\beta_{X}=0$ and $\gamma_{X}=0$).  We varied the following: the sample size, $n = (50, 100, 250, 500, 1000, 2000)$, the intercept, $\beta_{0} = (0.5, 1.0, 1.5, 2.0, 2.5)$, and the probability of a structural zero, $\omega = (0, 0.05, 0.1, 0.2, 0.5)$. We also varied the degree of overdispersion by setting  $\phi = \nu/\lambda = (\infty, 2, 1, 1/2, 1/3)$ (so that data simulated from the Negative Binomial distribution has a dispersion index of $d = 1 + \lambda/\nu = 1+1/\phi = (1.0, 1.5, 2.0, 3.0, 4.0)$).  To be clear, we consider:

\begin{itemize}
    \item scenarios with $\phi=\infty$ and $\omega=0$ as those with data simulated from the Poisson distribution;
    \item scenarios with $\phi<\infty$ and $\omega=0$ as those with data simulated from the Negative Binomial distribution;
    \item scenarios with $\phi=\infty$ and $\omega>0$ as those with data simulated from the Zero-inflated Poisson distribution; and
    \item scenarios with $\phi<\infty$ and $\omega>0$ as those with data simulated from the Zero-Inflated Negative Binomial distribution.
\end{itemize}    

We considered $X_i$ as a univariate continuous covariate from a Normal distribution, with mean of zero and variance of 100: $X_i \sim Normal (0, 100)$, for $i$ in 1,..., $n$ (as such, $k=2$).  Note that the covariate matrix $X$ is simulated anew for each individual simulation run.  Therefore, we are considering the case of \emph{random} regressors.   \citet{chen2011finite} discuss the difference between fixed and random covariates.  The assumption of fixed covariates is generally considered only in experimental settings whereas an assumption of random covariates is typically more appropriate for observational studies.


 \textcolor{black}{Note that, for the Poisson distributed data, we are simulating data with overall mean of $\lambda = exp(\beta_{0}) \approx (1.6,  2.7, 4.5, 7.4, 12.2)$.  For $\lambda>5$, zeros in the data are quite rare since $Pr(Y=0|\lambda) \approx 0$.  The simulation study could be expanded in several ways.  For instance, we did not consider models that deal with under-dispersion, even though under-dispersed counts may arise in various ecological studies; see \cite{lynch2014dealing}.  Also note that the simulation study only tests for rates of false-positives (since $\beta_{X}=0$ and $\gamma_{X}=0$ for all scenarios).  We are not testing for excessive false-negatives (and overly wide confidence intervals) which are also undesirable \citep{brooks2019statistical}. }
 

    
    
    

In total, we considered 750 distinct scenarios and for each simulated 15,000 unique datasets.  For each dataset, we conducted the seven step procedure and recorded all $p$-values and whether or not the null hypotheses is rejected at the 0.05 significance level under the entire procedure.  We also recorded all AIC statistics.  We are interested in the unconditional type one error.    We specifically chose to conduct 15,000 simulation runs so as to keep computing time within a reasonable limit while also reducing the amount of Monte Carlo standard error to a negligible amount (for looking at type 1 error with $\alpha=0.05$, Monte Carlo SE will be approximately $0.0018 \approx \sqrt{0.05(1-0.05)/15,000}$; see \cite{morris2019using}). 

To test the association between $X$ and $Y$ with each of the regression models, we conducted a Wald test to obtain the necessary $p$-value since in \verb|R|, the $p$-values in the default summary.glm output are from Wald tests.  Moreover, in initial simulations, LRTs performed rather erratically in rare situations when the model was misspecified (e.g., when a Poisson model was fit to ZIP data).

\section{Discussion}

\paragraph*{Analysis under the ``correct model'' - }

\begin{figure}
    \centering
    \includegraphics[width=14.5cm]{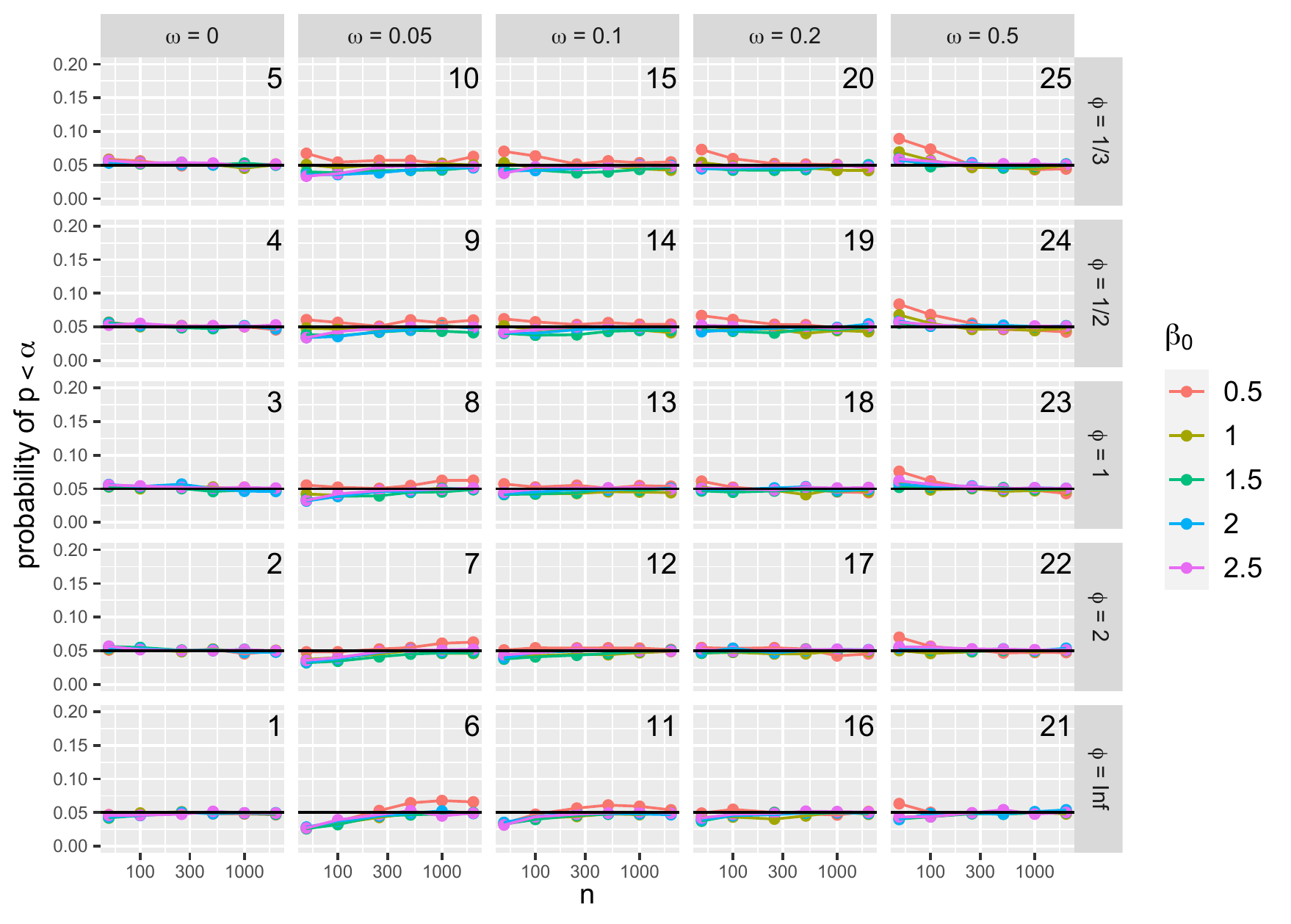}
    \caption{The empirical level of Type 1 error obtained under the ``correct'' model.  For panel 1, the ``correct'' model is the Poisson GLM; for panels 2-5 the ``correct'' model is the NB GLM;  for panels 6, 11, 16, 21, the ``correct'' model is the ZIP GLM; and for other panels, the ``correct'' model is the ZINB GLM.}
    \label{fig:correct}
\end{figure}

We first wish to confirm that the models under investigation deliver correct type 1 error when used as intended.  In other words, suppose the ``correct model'' is known a-\emph{priori} and  is used regardless of any preliminary diagnostic testing, would we obtain the desired 0.05 level of type 1 error?  See Figure \ref{fig:correct} which plots the rejection rates corresponding to this question.  

In summary, we see that for data simulated from the Poisson distribution (Figure \ref{fig:correct}, panel 1), empirical type 1 error is slightly smaller than 0.05 for small sample-size scenarios ($n \le 100$) and approximately 0.05 otherwise.  We also note that for data simulated from the NB distribution (Figure \ref{fig:correct}, panels 2-5), empirical type 1 error is approximately 0.05 for all $n \ge 100$ and for all $\beta_{0}$.  For data simulated from the ZIP distribution (see Figure \ref{fig:correct}, panels 6, 11, 16, 21), empirical type 1 error can be substantially conservative (i.e., less than 0.05) for small values of $n$ and small values of $\omega$.  Finally, for ZINB data, we note that, when $n$ is small,  the type 1 error appears to be higher than the advertised rate of 0.05 for some scenarios and less than 0.05 for others.  For example,  with $n = 100$, $\beta_{0}=0.5$, $\phi = 1/3$, and $\omega=0.5$, the type 1 error is 0.074, whereas, when  $n = 100$, $\beta_{0}=2.5$, $\phi = 2$, and $\omega=0.05$, the type 1 error is 0.040 (see Figure \ref{fig:correct}, panels 7 and 25).

We also note that none of the models appear to be ``robust'' to model misspecification.  The Poisson model applied to non-Poisson data leads to very high rejection rates (so high they are often off-the chart in Appendix - Figure \ref{fig:poisson}).  The ZIP model also performs poorly when applied to non-ZIP data (see Appendix - Figure \ref{fig:zip}), as does the NB model  when applied to non-NB data (see Appendix - Figure \ref{fig:nb}), and as does the ZINB model  (see Appendix - Figure \ref{fig:zinb}) when applied to non-ZINB data (specifically when applied to Poisson data and NB data).

As such, it seems inadvisable to recommend simply fitting a ZIP or ZINB to Poisson data if one is uncertain about the possibility of zero-inflation, or overdispersion.  As the sample size, $n$, increases (and as $\beta_{0}$ decreases), the type 1 error rates obtained when the ZIP and ZINB models are fit to Poisson data increase well beyond 0.05 (see Figures \ref{fig:zip} and \ref{fig:zinb}, panel 1).  This unexpected result may be due to the fact that these models are testing a null hypothesis that lies on the boundary of the parameter space (i.e., $\omega=0$).

\paragraph*{Preliminary testing - }

The next question is ``how often do the preliminary tests reject their null hypotheses?''  We also wish to determine how often the preliminary testing scheme successfully identifies the ``correct'' model.   

\begin{figure}
    \centering
    \includegraphics[width=14cm]{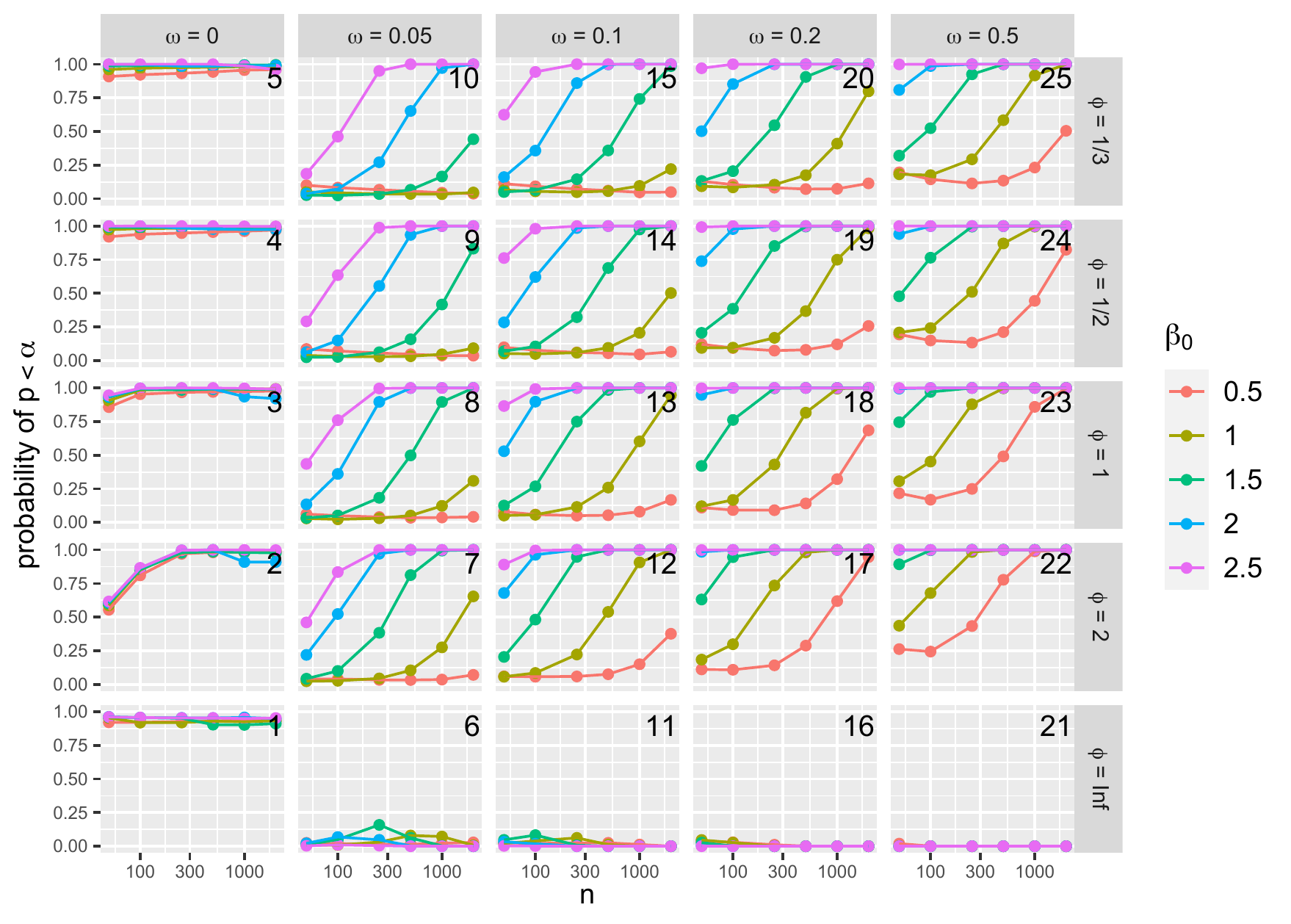}
    \caption{The probability of selecting the ``correct model'' after following the seven step testing scheme outlined in Section \ref{sec:methods}.}
    \label{fig:correct_model}
\end{figure}

Let us first consider the D\&L score test (see Appendix - Figure \ref{fig:overdispersion}) and specifically as it applies to the NB scenarios.  Recall that the NB scenarios are those with overdispersion ($\phi<\infty$) but no structural zero-inflation ($\omega=0$).  With the exception of the small sample-size scenarios with a small amount of overdispersion ($n\le100, \phi\le2$), the D\&L test correctly rejects the null hypothesis of no overdispersion for the vast majority of cases (Appendix - Figure \ref{fig:overdispersion}, panels 2-5).  For Poisson data (when $\phi=\infty$ and $\omega=0$), the D\&L test appears to show approximately correct type 1 error, with rejection rates ranging from 0.0368 to 0.0514 (see Figure \ref{fig:overdispersion}, panel 1).  However, for ZIP data (when $\phi=\infty$ and $\omega>0$), the D\&L test will often reject the null hypothesis of no overdispersion; see Figure \ref{fig:overdispersion}, panels 6, 11, 16, 21.  The rate of rejection increases with increasing sample size, with increasing $\omega$, and with increasing $\beta_{0}$.  Strictly speaking, rejection in these cases is correct since an excess of zeros ($\omega>0$) does contribute to overdispersion.  \textcolor{black}{However, it must be noted that using the NB model for overdispersion when the underlying issue is zero-inflation is not appropriate; see \citet{harrison2014using}.  Indeed, when the NB model is fit to ZIP data, we record type 1 error rates either much too low or much too high, depending on $\omega$, $\beta_{0}$, and $n$; see Figure \ref{fig:nb}, panels 6, 11, 16, and 21.}

Now let us discuss the Vuong test for zero-inflation.  See Appendix - Figures \ref{fig:vuongP} and \ref{fig:vuongNB} for the Vuong test results.  Note that the ``Poisson vs. ZIP'' Vuong test will often reject the null of no zero-inflation for NB data (Appendix - Figure \ref{fig:vuongP}, panels 2-5).  In contrast, the ``NB vs. ZINB'' Vuong test will rarely reject the null of no zero-inflation for NB data (Appendix - Figure \ref{fig:vuongNB}, panels 2-5).  In this way, the Vuong test acts as a second-line defense against erroneously selecting the Poisson model.  If the D\&L score test fails to select the NB model in Step 1, the ``Poisson vs. ZIP'' Vuong test in Step 3  will often reject the Poisson model in favour of the ZIP model (particularly when $n$ and $\beta_{0}$ are large).  The ZIP model, when used for NB, is not ideal, but is definitely preferable to the Poisson model; compare Appendix - Figures \ref{fig:poisson} and \ref{fig:zip}, panels 2-5.  

Overall, the probability that the preliminary seven-step testing scheme selects the ``correct'' model depends highly on $\beta_{0}$, $\omega$, $\phi$, and $n$, see Figure \ref{fig:correct_model}.  With Poisson data, if each of the diagnostic tests were truly  independent (and each had a $\alpha=0.05$ type 1 error rate), then the probability of selecting the ``correct'' model should be 90.25\% ($=0.95\times95\%$); see Figure \ref{fig:tree}.  The numbers we obtain from the simulation study range from 90.21\% to 96.19\%.

For the ZIP data scenarios ($\omega>0$, $\phi=\infty$; Figure \ref{fig:correct_model}, panels 6, 11, 16, 21), the ``incorrect'' ZINB model is chosen in a majority of cases.  This may not necessarily lead to type 1 error inflation since the ``incorrect'' ZINB model is often conservative when applied to ZIP data; see  Appendix - Figure \ref{fig:zinb}.  For ZINB data scenarios (i.e., when $\omega>0$ and $\phi<\infty$), in cases when the ZINB model is not selected, it is most likely that the NB model is selected instead.  This also might not necessarily lead to type 1 error inflation since the misspecified NB model appears to maintain a type 1 error rate at or bellow the advertised rate in many of these situations (specifically when $\phi<2$ and $\omega<0.2$); see Appendix - Figure \ref{fig:nb}.  

\paragraph*{Post-testing unconditional type 1 error - }

\begin{figure}[h!]
    \centering
\includegraphics[width=14cm]{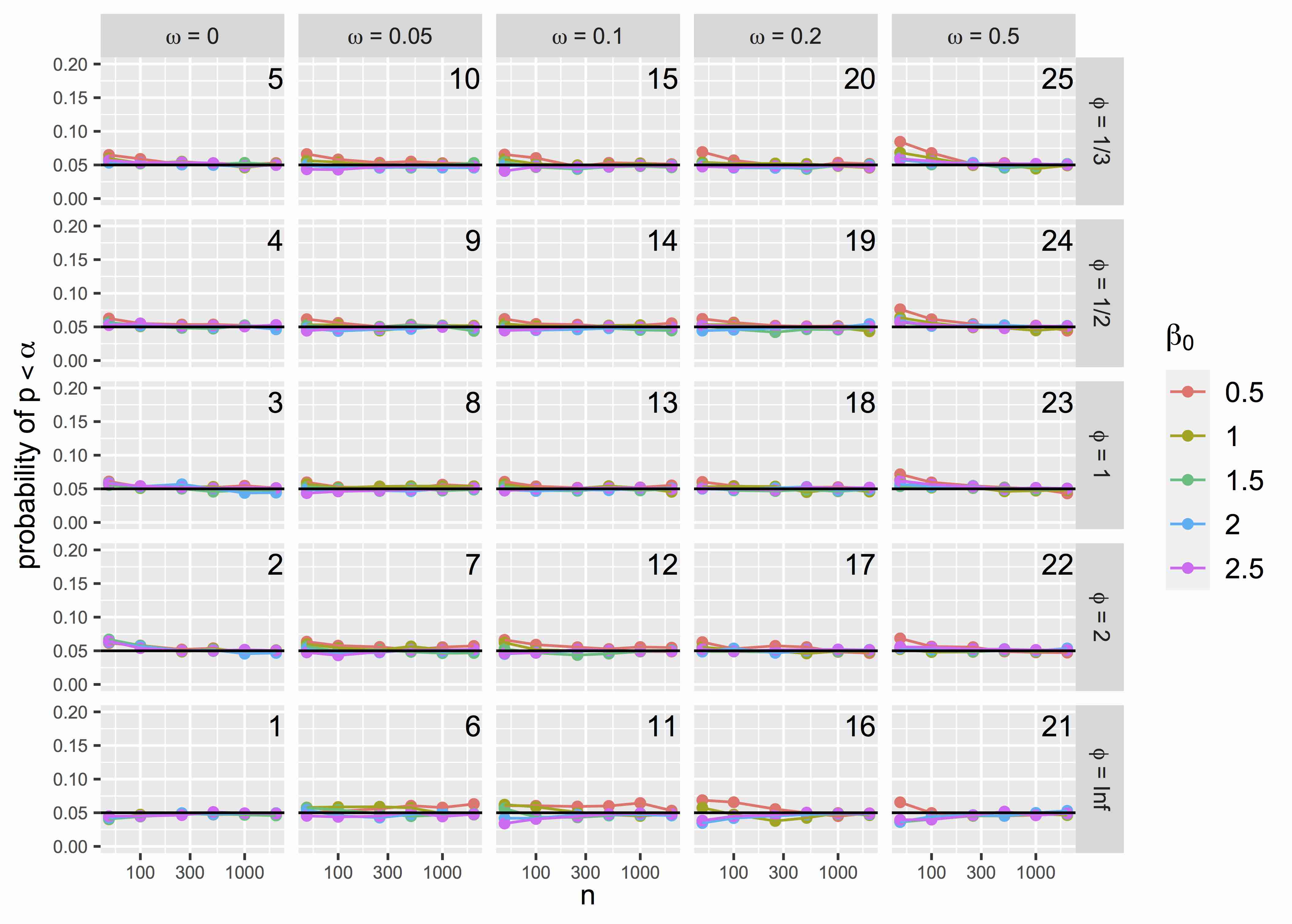}
    \caption{Type 1 error obtained following the seven step testing scheme outlined in Section \ref{sec:methods}.}
    \label{fig:type1}
\end{figure}

Our main question of interest is whether or not the null hypotheses of no association between $X$ and $Y$ is rejected at the desired 0.05 significance level when following the entire seven-step procedure  outlined in Section \ref{sec:methods}.  The corresponding rejection rates are plotted in Figure \ref{fig:type1}.  Table \ref{tab:rates} lists rejection rates and model selection rates for a select number of scenarios.  Let us consider the results for each distribution.

\begin{table}[h!]
    \centering
    \begin{footnotesize}
    \begin{tabular}{|lcccc|}
     \hline
     &  \textbf{Poisson GLM}  & \textbf{ZIP GLM} & \textbf{NB GLM} & \textbf{ZINB GLM} \\
    \hline

\multicolumn{3}{|l}{\textbf{Scenario ``3''}} & &\\       
\multicolumn{3}{|l}{($n=250$, $\beta_{0}=0.5$, $\phi = \infty$, $\omega= 0$; Poisson)} & &\\       
   \hspace{0.1cm}  $\operatorname{Pr}( \textrm{reject $H_{0}$}  )$&  0.05 & 0.04 & 0.05 & 0.04 \\ 
        \hspace{0.1cm}  $\operatorname{Pr}(\textrm{reject $H_{0}$} | \textrm{$M$ selected by tests})$&  0.05 & 0.04 & 0.02 & 0.00 \\ 
     \hspace{0.1cm} $\operatorname{Pr} (\textrm{$M$ selected by tests})$& 0.93 & 0.02 & 0.05 & 0.00 \\ 
              \hspace{0.1cm}  $\operatorname{Pr}(\textrm{reject $H_{0}$} | \textrm{$M$ has lowest AIC})$&  0.05 & 0.16 & 0.02 & 0.04 \\ 
          \hspace{0.1cm} $\operatorname{Pr} (\textrm{$M$ has lowest AIC})$&  0.83 & 0.11 & 0.05 & 0.01 \\
          
\multicolumn{3}{|l}{\textbf{Scenario ``6''}} & &\\       
\multicolumn{3}{|l}{($n=2,000$, $\beta_{0}=0.5$, $\phi = \infty$, $\omega= 0$; Poisson)} & &\\       
   \hspace{0.1cm}  $\operatorname{Pr}( \textrm{reject $H_{0}$}  )$&    0.05 & 0.07 & 0.05 & 0.06 \\ 
 \hspace{0.1cm}  $\operatorname{Pr}(\textrm{reject $H_{0}$} | \textrm{$M$ selected by tests})$& 0.05 & 0.09 & 0.04 & 0.30 \\ 
 \hspace{0.1cm} $\operatorname{Pr} (\textrm{$M$ selected by tests})$& 0.93 & 0.02 & 0.05 & 0.00 \\ 
\hspace{0.1cm}  $\operatorname{Pr}(\textrm{reject $H_{0}$} | \textrm{$M$ has lowest AIC})$&  0.05 & 0.30 & 0.04 & 0.13 \\ 
          \hspace{0.1cm} $\operatorname{Pr} (\textrm{$M$ has lowest AIC})$&  0.83 & 0.10 & 0.06 & 0.01 \\     
          
\multicolumn{3}{|l}{\textbf{Scenario ``36''}} & &\\       
\multicolumn{3}{|l}{($n=2,000$, $\beta_{0}=0.5$, $\phi = 2$, $\omega= 0$; NB)} & &\\            
   \hspace{0.1cm}  $\operatorname{Pr}( \textrm{reject $H_{0}$}  )$& 0.11 & 0.08 & 0.05 & 0.07 \\ 
 \hspace{0.1cm}  $\operatorname{Pr}(\textrm{reject $H_{0}$} | \textrm{$M$ selected by tests})$&  -- & --  &  0.05 & 0.12 \\ 
\hspace{0.1cm} $\operatorname{Pr} (\textrm{$M$ selected by tests})$&   0.00 & 0.00 & 0.98 & 0.02 \\ 
 \hspace{0.1cm}  $\operatorname{Pr}(\textrm{reject $H_{0}$} | \textrm{$M$ has lowest AIC})$& --  & -- & 0.05 & 0.24 \\ 
           \hspace{0.1cm} $\operatorname{Pr} (\textrm{$M$ has lowest AIC})$&  0.00 &  0.00 & 0.87 & 0.13 \\           
    
\multicolumn{3}{|l}{\textbf{Scenario ``43''}} & &\\       
\multicolumn{3}{|l}{($n=50$, $\beta_{0}=1.5$, $\phi = 2$, $\omega= 0$; NB)} & &\\       
   \hspace{0.1cm}  $\operatorname{Pr}( \textrm{reject $H_{0}$}  )$&  0.10 & 0.04 & 0.06 & 0.02 \\ 
 \hspace{0.1cm}  $\operatorname{Pr}(\textrm{reject $H_{0}$} | \textrm{$M$ selected by tests})$&  0.11 & 0.00 & 0.04 & 0.18 \\ 
\hspace{0.1cm} $\operatorname{Pr} (\textrm{$M$ selected by tests})$&  0.40 & 0.00 & 0.60 & 0.00 \\ 
 \hspace{0.1cm}  $\operatorname{Pr}(\textrm{reject $H_{0}$} | \textrm{$M$ has lowest AIC})$&  0.10 & 0.09 & 0.04 & 0.03 \\ 
           \hspace{0.1cm} $\operatorname{Pr} (\textrm{$M$ has lowest AIC})$& 0.33 & 0.08 & 0.56 & 0.03 \\ 
    
     \multicolumn{3}{|l}{\textbf{Scenario ``182''}} & &\\
    \multicolumn{3}{|l}{($n=100$, $\beta_{0}=0.5$, $\phi = 2$, $\omega= 0.05$; ZINB)} & &\\
        \hspace{0.1cm}  $\operatorname{Pr}( \textrm{reject $H_{0}$}  )$& 0.12 & 0.08 & 0.05 & 0.05 \\ 
        \hspace{0.1cm}  $\operatorname{Pr}(\textrm{reject $H_{0}$} | \textrm{$M$ selected by tests})$  &   0.12 & 0.29 & 0.05 & 0.14 \\ 
     \hspace{0.1cm} $\operatorname{Pr} (\textrm{$M$ selected by tests})$ & 0.09 & 0.00 & 0.87 & 0.04 \\ 
             \hspace{0.1cm}  $\operatorname{Pr}(\textrm{reject $H_{0}$} | \textrm{$M$ has lowest AIC})$ &   0.12 & 0.13 & 0.05 & 0.14 \\ 
          \hspace{0.1cm} $\operatorname{Pr} (\textrm{$M$ has lowest AIC})$ &  0.06 & 0.19 & 0.67 & 0.08 \\

     \multicolumn{3}{|l}{\textbf{Scenario ``302''}} & &\\
   \multicolumn{3}{|l}{ ($n=100$, $\beta_{0}=0.5$, $\phi = \infty$, $\omega= 0.1$; ZIP)} & &\\     
    \hspace{0.1cm} $\operatorname{Pr}( \textrm{reject $H_{0}$} )$& 0.06 & 0.05 & 0.05 & 0.04 \\ 
    \hspace{0.1cm} $\operatorname{Pr}(\textrm{reject $H_{0}$} | \textrm{$M$ selected by tests})$  & 0.07 & 0.17 & 0.03 & 0.15 \\
     \hspace{0.1cm} $\operatorname{Pr} (\textrm{$M$ selected by tests})$ & 0.71 & 0.02 & 0.25 & 0.02 \\ 
         \hspace{0.1cm} $\operatorname{Pr}(\textrm{reject $H_{0}$} | \textrm{$M$ has lowest AIC})$ &  0.06 & 0.10 & 0.03 & 0.07 \\
          \hspace{0.1cm} $\operatorname{Pr} (\textrm{$M$ has lowest AIC})$ & 0.51 & 0.31 & 0.16 & 0.02 \\ 
     
        \hline
    \end{tabular}
    \end{footnotesize}
    \caption{Rejection rates and model selection rates for a number of selected scenarios from the simulation study. These numbers can be used to calculate the overall unconditional type 1 error rates.  For example,  for Scenario ``302'', the  type 1 error obtained after model selection via AIC is 0.071 (=$ 0.06\times0.51 +  0.10\times0.31 + 0.03\times0.16 + 0.07\times0.02$);  and the  type 1 error obtained after model selection via sequential score tests is  0.060 (=$0.07\times0.71 + 0.17\times0.02 + 0.03\times0.25 + 0.15\times0.02$).}
    \label{tab:rates}
\end{table}

First, for data simulated from the Poisson distribution (Figure \ref{fig:type1}, panel 1), empirical type 1 error appears to be unaffected by model selection bias.  This is due to the fact that incorrect models are rarely selected, even when sample sizes are small (see  Figure \ref{fig:correct_model}, panel 1).  Consider two specific scenarios: 

\begin{itemize}
    \item  Scenario ``3'' ($n=250$, $\beta_{0}=0.5$, $\phi=\infty$, and $\omega=0$) - When $\beta_{0}=0.5$ and $n=250$, the Poisson model is correctly selected in approximately 93\% of cases while the NB and ZIP models are selected in about 5\% and 2\% of cases, respectively. Numbers in the top right-hand corner of each node in Figure \ref{fig:tree} indicate the expected number of datasets (out of a total of 100) to reach each outcome if the data was Poisson (with $\beta_{X}=0$), and each of the tests were truly independent (with a $\alpha=0.05$ type 1 error rate).  The numbers in parentheses correspond to results from the simulation study for this scenario.
    
    For those datasets directed to the NB and ZIP models, the null hypothesis of no association between $X$ and $Y$ is rejected with probability of 0.021 (=0.10/(4.63+0.10)) and 0.044 (=0.11/(2.34+0.11)), respectively.  As such, model selection bias, in this case, has the innocuous effect of ever so slightly lowering the type 1 error level: the Poisson GLM fit to this data provides a type 1 error rate of 0.050, whereas the unconditional type 1 error rate obtained after following the seven-step procedure is 0.049 (=0.0471+0.0011+0.0010+0.0000).
    
    \item Scenario ``6'' ($n=2,000$, $\beta_{0}=0.5$, $\phi=\infty$, and $\omega=0$) - When $\beta_{0}=0.5$ and $n=2,000$, the Poisson model is correctly selected in approximately 93\% of cases while the NB and ZIP models are selected in about 5\%  and 2\% of cases, respectively.  While the NB model is conservative for this data ($\operatorname{Pr}(\textrm{reject $H_{0}$} | \textrm{NB model selected by tests}) = 0.037$), the ZIP model is not ($\operatorname{Pr}(\textrm{reject $H_{0}$} | \textrm{ZIP model selected by tests}) = 0.089$).  However, the impact is negligible: the unconditional type 1 error rate obtained after following the seven-step procedure is 0.047.
    

\end{itemize}

Second, for data simulated from the ZIP distribution (i.e., when $\omega>0$ and $\phi=\infty$), the ``incorrect'' ZINB model is almost always selected due to the fact that the model selection procedure tests for zero-inflation only after first testing for overdispersion.  However, the type 1 error under this ``incorrect'' ZINB model is, for most scenarios, not substantially higher than the advertised 0.05 rate, (see Appendix - Figure \ref{fig:zinb}, panels 6, 11, 16, 21).  There are, however, exceptions where model selection bias is apparent.  Consider, for example, scenario ``302'': 

\begin{itemize}
    \item Scenario ``302'' ($n=100$, $\beta_{0}=0.5$, $\phi = \infty$ and  $\omega= 0.1$) - The unconditional type 1 error obtained after following the seven step procedure is 0.060 (see Figure \ref{fig:type1}, panel 11). Amongst the simulated datasets for which the ZIP model is selected (by the D\&L and Vuong tests), the ZIP model has a rejection rate of 0.168.  Amongst the simulated datasets for which the ZINB model is selected, the ZINB model has a rejection rate of 0.154; see Table \ref{tab:rates}.  This clearly shows that the diagnostic tests (the D\&L and Vuong tests) and the subsequent hypothesis tests ($H_{0}: \beta_{X}=\gamma_{X}=0$) are not independent of one another.  In this instance, the D\&L test will not only screen for overdispersion, but will also direct the data towards a model that is more likely to reject $H_{0}: \beta_{X}=\gamma_{X}=0$, thereby inflating the type 1 error.
\end{itemize}



With data simulated from the NB distribution (i.e., when $\phi<\infty$ and $\omega=0$; see Figure \ref{fig:type1}, panels 2-5), we see that model selection bias can lead to modest type 1 error inflation when $n$ is small.   When sample sizes are sufficiently large, there is little evidence of any substantial type 1 error inflation caused by model selection bias.  Consider for example ``Scenario 43'':

\begin{itemize}
    \item Scenario ``43'' ($n=50$, $\beta_{0}=1.5$, $\phi = 2$ and  $\omega= 0$) - The unconditional type 1 error obtained after following the seven step procedure is 0.067 (see Figure \ref{fig:type1}, panel 2), whereas the type 1 error obtained with the ``correct'' NB model is 0.056.  This inflation is due to the fact that, for this data, there is a 40\% probability of selecting the Poisson model following the seven-step procedure and that  $\operatorname{Pr}(\textrm{reject $H_{0}$} | \textrm{Poisson model is selected by tests})=0.11$; see Table \ref{tab:rates}.
\end{itemize}

Finally, consider data simulated from the ZINB distribution (i.e., when $\phi<\infty$ and $\omega>0$; see Figure \ref{fig:type1}, panels 7-10, 12-15, 17-20, 22-25).  We see type 1 error rates much higher than 0.05 for some scenarios (e.g., when $n$ is small and $\omega$ is large).   For example, consider scenario ``182'':

\begin{itemize}
    \item Scenario ``182'' ($n=100$, $\beta_{0}=0.5$, $\phi = 2$ and  $\omega= 0.05$) - The unconditional type 1 error obtained after following the seven step procedure is 0.058 (see Figure \ref{fig:type1}, panel 7), whereas the type 1 error obtained with the ``correct'' ZINB model is 0.048.  Note that, when applied to the data ignoring the results of the diagnostic tests, both the NB and the ZINB models demonstrate appropriate rejection rates (of 0.051 and 0.048, respectively; see Table \ref{tab:rates}).  However, amongst the simulated datasets for which the ZINB model is selected (by the D\&L and Vuong diagnostic tests) the ZINB model has a rejection rate of 0.142.  This clearly shows that, to the detriment of the type 1 error rate, the diagnostic tests and the subsequent hypothesis test for $H_{0}: \beta_{X}=\gamma_{X}=0$ are not independent.
\end{itemize}





\paragraph*{AIC model selection - }

\begin{figure}[h!]
    \centering
\includegraphics[width=14cm]{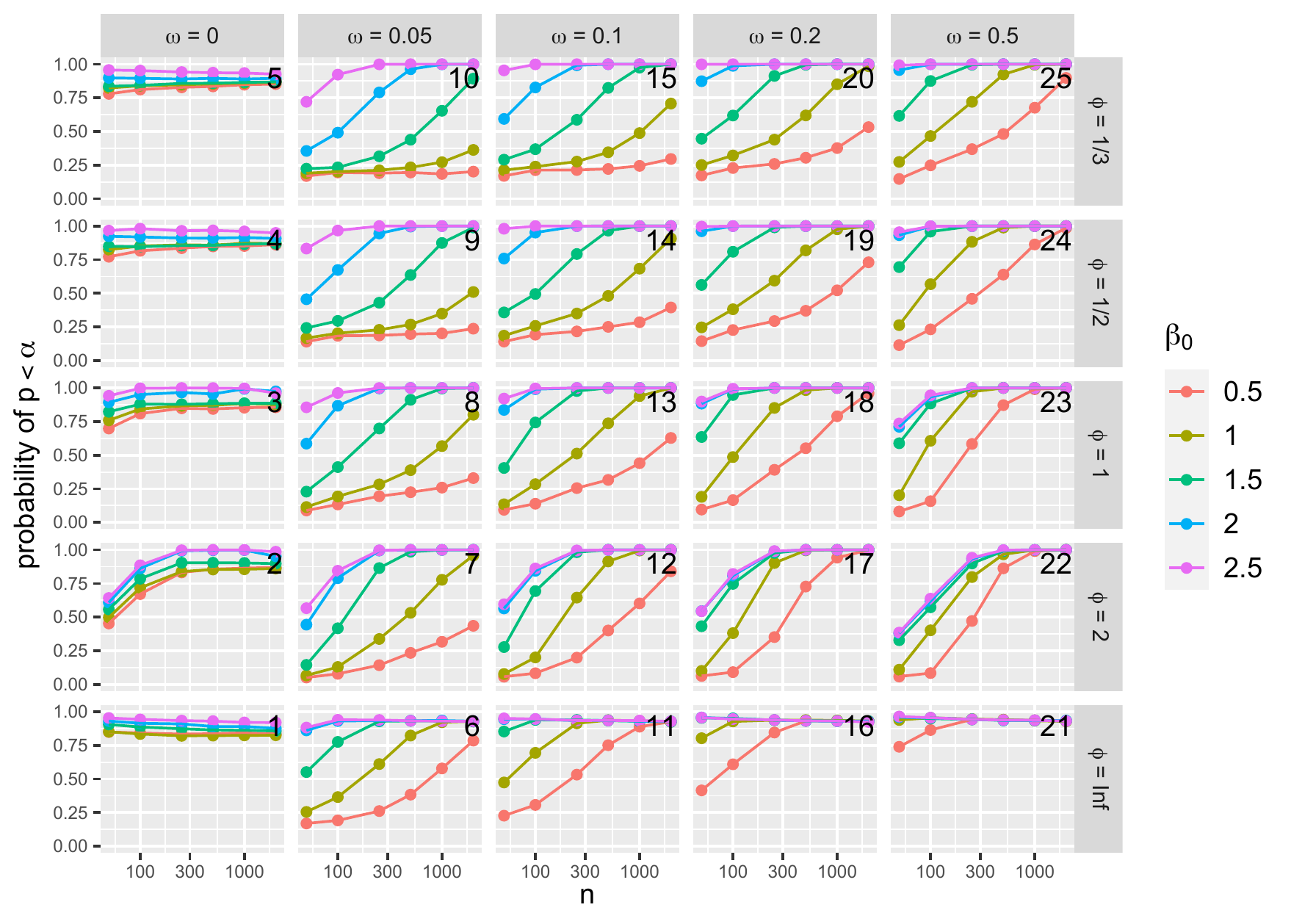}
    \caption{The probability that the ``correct'' model is the one with the lowest AIC.}
    \label{fig:correct_selectionAIC}
\end{figure}

We also investigated model selection using the AIC.  We were curious as to how often the ``correct'' model is the model with the lowest AIC. Figure \ref{fig:correct_selectionAIC} plots the results.  We see that the probability that the AIC statistic selects the ``correct'' model depends highly on $\beta_{0}$, $\omega$, $\phi$, and $n$.  Overall, across all scenarios we considered, the AIC selected the correct model for 77\% of datasets whereas the seven-step model selection based on score tests selects the correct model for 58\% of datasets.

\begin{figure}[h!]
    \centering
\includegraphics[width=14cm]{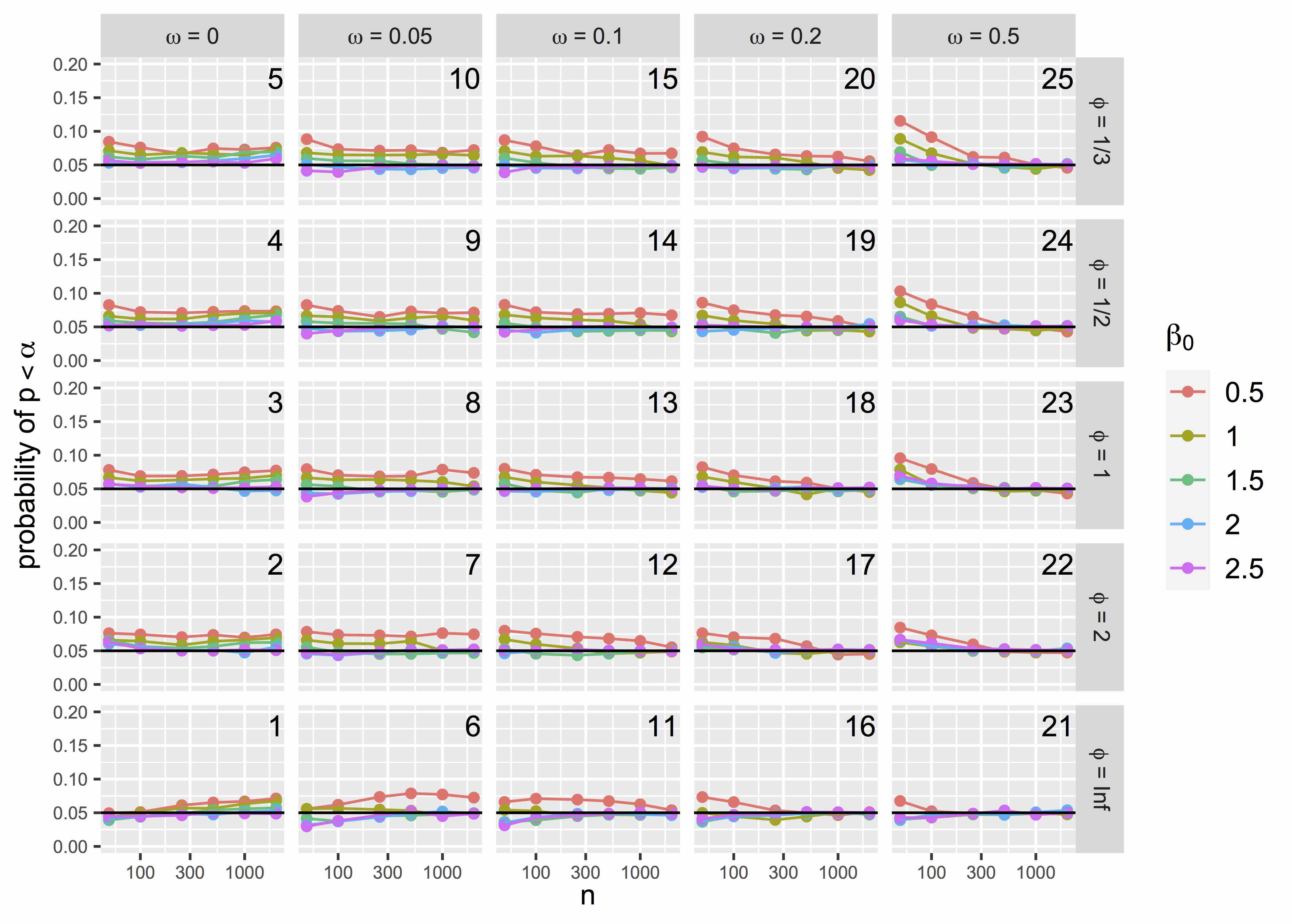}
    \caption{Type 1 error obtained from model with the lowest AIC.}
    \label{fig:type1AIC}
\end{figure}

More specifically, for the NB data scenarios ($\omega=0$, $\phi<\infty$; Figure \ref{fig:correct_selectionAIC}, panels 2-5), the ``correct'' NB model is chosen using the AIC in a large majority of cases for most scenarios.  In contrast, for ZIP data (i.e., when $\omega>0$ and $\phi=\infty$), the AIC is less capable of determining the ``correct model'' when $\beta_{0}$, $n$, and $\omega$ are small.  For ZINB data scenarios (when $\omega>0$ and $\phi<\infty$), the probability of selecting the correct model using the AIC ranges substantially and increases  (somewhat predictably) with increasing $n$ and increasing $\beta_{0}$.

We also wish to determine whether or not the null hypotheses of no association between $X$ and $Y$ is rejected at the 0.05 significance level when following model selection via AIC.  Figure \ref{fig:type1AIC} shows that, when $\beta_{0}$ is small, there are several scenarios in which the unconditional type 1 error is much too high.  Perhaps most surprisingly, with Poisson data (i.e., scenarios with $\omega=0$ and $\phi=\infty$), when $\beta_{0}=0.5$, the unconditional type 1 error increases with increasing $n$, from  0.049 to 0.071 (see Figure \ref{fig:type1AIC}, panel 1).    Consider again Scenario ``3'' ($n=250$, $\beta_{0}=0.5$, $\phi=\infty$, and $\omega=0$) and Scenario ``6'' ($n=2,000$, $\beta_{0}=0.5$, $\phi=\infty$, and $\omega=0$); see Table \ref{tab:rates}. 

\begin{itemize}
    \item For Scenario ``3'', the probability that the AIC incorrectly selects the ZIP GLM is 0.11, and $\operatorname{Pr} (\textrm{reject $H_{0}$}|\textrm{ZIP GLM has lowest AIC}) = 0.16$.  The unconditional type 1 error rate = 0.06 (=$0.05\times0.83 + 0.16\times0.11 +  0.02\times0.05 + 0.04\times0.01$).
    
    \item  For Scenario ``6'', the probability that the AIC incorrectly selects the ZIP GLM is 0.10, and $\operatorname{Pr} (\textrm{reject $H_{0}$}|\textrm{ZIP GLM has lowest AIC}) = 0.30$.  The unconditional type 1 error rate = 0.07 ($= 0.05\times0.83 + 0.30\times0.10 + 0.04\times0.06 + 0.13\times0.01$).
\end{itemize}

 With NB data (i.e., scenarios with $\omega=0$ and $\phi<\infty$), the unconditional type 1 error is also much higher than 0.05, even when $n$ and $\beta_{0}$ are large.   This is due to the fact that the ZINB model, when erroneously selected in a minority of cases, rejects the null of no association between $X$ and $Y$ at rates much much higher than 0.05.  This particularly true when $n$ is large.  Consider for example, Scenario ``36'':
 
 \begin{itemize}
 \item
 
  Scenario ``36'' ($n=2,000$, $\beta_{0}= 0.5$,  $\phi=2$, and $\omega= 0$) -   Amongst the 87\% of datasets for which the AIC correctly selects the NB model, the null hypothesis of no association between $X$ and $Y$ is rejected  with probability of exactly 0.050; see Table \ref{tab:rates}.  However, amongst the remaining 13\% of datasets for which the ZINB model is erroneously selected, the probability of rejecting the null hypothesis of no association between $X$ and $Y$ is 0.240.  As a result the unconditional type 1 error rate is 0.074 ($=0.240\times0.13 + 0.87\times0.05$).
\end{itemize}

In summary, while the AIC is able to select the ``correct'' model more often than the sequential score testing scheme, there appears to be more potential for type 1 error inflation.  How can this be?  In the presence of model selection bias, selecting the ``correct'' model more often is, somewhat paradoxically, not always preferable.  Consider once again Scenario ``302'' (with $n=100$, $\beta_{0}=0.5$, $\phi = \infty$ and  $\omega= 0.1$).  Following  the sequential score tests, the ``correct'' ZIP model was only selected with a 2\% probability.  With model selection via AIC, the ``correct'' ZIP model was selected with a 31\% probability.    However, the unconditional type 1 error obtained after following the seven step procedure is 0.060, whereas the unconditional type 1 error obtained after model selection via AIC is 0.071; see Table \ref{tab:rates}.  We see a similar phenomenon with Scenario ``182'' (with $n=100$, $\beta_{0}=0.5$, $\phi = 2$ and  $\omega= 0.05$); see Table \ref{tab:rates}.  The ``correct'' ZINB model is chosen more often with the AIC than with the score tests (8\%  vs. 4\%).  However,  the  type 1 error obtained after model selection via AIC is 0.074, vs. 0.058 after model selection by sequential score tests.

\section{Conclusions}

If the population distribution is known in advance, model selection bias will not be a problem. If the assumptions required of the Poisson distribution are known to be wrong, alternative models that do not depend on these assumptions can be used and ideally a valid model can be pre-specified prior to obtaining/observing any data.  However, outside of a highly controlled laboratory experiment, this may not be realistic.  The potentially problematic (and most likely scenario) is when one cannot, with a high degree of confidence, determine the distributional nature of the data before observing the data.  What should be done in these circumstances?  \cite{tsou2006robust} suggest using a ``robust'' Poisson regression model ``so that one need not worry about the correctness of the Poisson assumption.''  However, when the distributional assumptions of the Poisson GLM do hold, \cite{tsou2006robust} acknowledge that the ``robust approach might not be as efficient.'' Given the potentially immense expense required to obtain data, anyone working in data-driven research will no doubt be reluctant to adopt any approach which compromises statistical power.

Researchers who do not know in advance whether or not there is overdispersion or zero-inflation, might decide to simply use a ZIP or ZINB as a ``safer bet'' \citep{perumean2013zero} and pay a price in terms of efficiency \citep{williamson2007power}.  However, this is problematic.  We observed that the ZIP and ZINB models, when fit to ordinary Poisson data, can lead to type 1 error well above the advertised rate when sample sizes are large.  (Future work should consider whether hurdle models \citep{rose2006use} are similarly problematic.)  Instead, if there is sufficient data, researchers should proceed with a model selection procedure, ideally one based on efficient score tests.  

Our simulation study suggests that, if sample sizes are sufficiently large, there is little need to worry about model selection bias following a series of sequential score tests. However, when sample sizes are small, our simulation study demonstrated that model selection bias can lead to potentially substantial type 1 error inflation.  

Model selection based on the AIC cannot be recommended.  We observed that, even when sample sizes are large, when the true underlying distribution of the data is Poisson, using the AIC to select the ``best'' model can often lead to substantial type 1 error inflation.  Future work should investigate the suitability of other information criteria (e.g., BIC, AICc).



Ignoring the possibility of overdispersion and zero inflation during data analyses can lead to invalid inference.  However, if one does not have sufficient power to confidently test for overdispersion and zero inflation, it \textcolor{black}{may be} best to simply use a model that can accommodate for these possibilities (e.g., use a robust model) instead of going through a model selection procedure that might inflate the type 1 error.  In summary, if one does not have the power to test for distributional assumptions, testing for distributional assumptions may not be wise.  And if one does have a sufficiently large sample size to test for distributional assumptions, testing for distributional assumptions may be very beneficial.  Note that our simulation study did not include any covariates and in studies where there are several covariates, it will no doubt be difficult to determine what constitutes a ``sufficiently large'' sample size.  To conclude, be reminded that researchers should always be cautious when interpreting results when $n$ is small \citep{button2013power}.  Model selection bias is just one more reason to have a healthy skepticism of NHST when small sample sizes are small.




\bibliography{truthinscience}


\vspace{3cm}

\section{Appendix}

Let us briefly review the \cite{dean1989tests}  score test for overdispersion and the Vuong test for zero-inflation.

 \noindent \textbf{(1) The $D\&L$ score test - } \citet{dean1989tests} proposed calculating the following score statistic for testing overdispersion:

\begin{equation}
T_{1}=\sum_{i=1}^{n}\left\{\left(y_{i}-\hat{\lambda}_{i}\right)^{2}-y_{i}\right\} /\left(2 \sum_{i=1}^{n} \hat{\lambda}_{i}^{2}\right)^{1 / 2}
\end{equation}

\noindent Under the null hypothesis of no overdispersion, the $T_{1}$ statistic converges to a standard Normal distribution and the $p$-value is calculated as: $p\textrm{-value} = P_{\mathcal{N}}(T_{1}).$

 \noindent \textbf{(2) The Vuong test for zero-inflation - }  The  Vuong test statistic is calculated as follows:

\begin{equation}
V = \frac{\sum_{i=1}^{n}(\textrm{log}dL_{i})}{\sqrt{n}\cdot \sqrt{\sum_{i=1}^{n}((\textrm{log}dL_{i}-\sum_{i=1}^{n}(\textrm{log}dL_{i})/n)^2/(n-1))}},
\end{equation}

\noindent where, if the Poisson model is compared to it's zero-inflated counterpart, the ZIP model, we define: $\textrm{log}dL_{i} = \textrm{log}(Pr_{ZIP}(Y_{i}=y_{i}|\widehat{\omega}_{i}, \widehat{\lambda}_{i})) - \textrm{log}(Pr_{Pois}(Y_{i}=y_{i}|\widehat{\lambda}_{i})).$ If the NB model is compared to the ZINB model, we define: $
    \textrm{log}dL_{i} = \textrm{log}(Pr_{ZINB}(Y_{i}=y_{i}|\widehat{\nu}, \widehat{\omega}_{i}, \widehat{\lambda}_{i})) - \textrm{log}(Pr_{NB}(Y_{i}=y_{i}|\widehat{\nu}, \widehat{\lambda}_{i})).$

The $V$ statistic, under the null, will follow the Normal distribution and a $p$-value is calculated as: $\textrm{$p$-value} = 1 - P_{N}(|V|)$.  Note that \citet{desmarais2013testing} have suggested an adjustment to the Vuong test which, for larger samples, may have greater efficiency.  Also, note that the Vuong test for zero-inflation, while widely used in practice, is somewhat controversial, see \citet{wilson2015misuse}.

\section{R-Code: Models and score tests}
\begin{footnotesize}
\begin{verbatim}

library("pscl", "lmtest")
#####################
### MODELS:

#####################
### Poisson model ###
PoissonGLM <- glm(y ~ newX, family = poisson)
Poisson_pval <- waldtest(PoissonGLM)[ "Pr(>F)"][2,]
Poisson_AIC <- AIC(PoissonGLM)

#####################
### NB model ###
NB_AIC<-Inf
NB_pval <- 0.99
	tryCatch({
		NB_mod  <- glm.nb(y ~ newX)
		NB_pval <- waldtest(NB_mod)[ "Pr(>F)"][2,]
		NB_AIC  <- AIC(NB_mod)
		}, 
		 error=function(e){})

#####################
### ZIP model ###
ZIP_AIC <- Inf
zip_mod <- 99
ZIP_pval <- 0.99
zip_modNA <- TRUE

tryCatch({
		zip_mod   <- zeroinfl(y ~ newX|newX, dist = "poiss")
		zip_modNA <- sum(is.na(unlist((summary(zip_mod))$coefficients)))>0		
		}, 
		error=function(e){})

	if(is.double(zip_mod)| zip_modNA){	
		tryCatch({		
		zip_mod <- zeroinfl(y ~ newX|newX , dist = "poiss", EM=TRUE)}, 
		error=function(e){})}
					
tryCatch({			
ZIP_pval <- waldtest(zip_mod)[ "Pr(>Chisq)"][2,]
ZIP_AIC <- AIC(zip_mod)
	 },
		 error=function(e){})

#####################
### ZINB model ###		 
ZINP_AIC 	<- Inf	 
zinb_mod<-99
ZINB_pval <- 0.99
zinb_modNA <- TRUE

tryCatch({
		zinb_mod <- zeroinfl(y ~ newX|newX , dist = "negbin")
		zinb_modNA <- sum(is.na(unlist((summary(zinb_mod))$coefficients)))>0
		}, 
		error=function(e){})
		 
		if(is.double(zinb_mod) | zinb_modNA){
		tryCatch({
		zinb_mod <- zeroinfl(y ~ newX|newX  , dist = "negbin", EM=TRUE); summary(zinb_mod)
		}, 
		 error=function(e){})}
		 

tryCatch({	
ZINB_pval<-waldtest(zinb_mod)[ "Pr(>Chisq)"][2,]
dimK<-dim(summary(zinb_mod)$coefficients$count)[1]
ZINP_AIC <-AIC(zinb_mod)
		},
		error=function(e){})

#####################
### SCORE TESTS:

#######
## vuong_f test :  compare model1 to model2 (modified from "pscl" package)
#######

vuong_f<-function (m1, m2, digits = getOption("digits")) 
{
    m1y <- m1$y
    m2y <- m2$y
    m1n <- length(m1y)
    m2n <- length(m2y)
    if (m1n == 0 | m2n == 0) 
        stop("Could not extract dependent variables from models.")
    if (m1n != m2n) 
        stop(paste("Models appear to have different numbers of observations.\n", 
            "Model 1 has ", m1n, " observations.\n", "Model 2 has ", 
            m2n, " observations.\n", sep = ""))
    if (any(m1y != m2y)) {
        stop(paste("Models appear to have different values on dependent variables.\n"))
    }
    p1 <- predprob(m1)
    p2 <- predprob(m2)
    if (!all(colnames(p1) == colnames(p2))) {
        stop("Models appear to have different values on dependent variables.\n")
    }
    whichCol <- match(m1y, colnames(p1))
    whichCol2 <- match(m2y, colnames(p2))
    if (!all(whichCol == whichCol2)) {
        stop("Models appear to have different values on dependent variables.\n")
    }
    m1p <- rep(NA, m1n)
    m2p <- rep(NA, m2n)
    for (i in 1:m1n) {
        m1p[i] <- p1[i, whichCol[i]]
        m2p[i] <- p2[i, whichCol[i]]
    }
    k1 <- length(coef(m1))
    k2 <- length(coef(m2))
    lm1p <- log(m1p)
    lm2p <- log(m2p)
    m <- lm1p - lm2p
    bad1 <- is.na(lm1p) | is.nan(lm1p) | is.infinite(lm1p)
    bad2 <- is.na(lm2p) | is.nan(lm2p) | is.infinite(lm2p)
    bad3 <- is.na(m) | is.nan(m) | is.infinite(m)
    bad <- bad1 | bad2 | bad3
    neff <- sum(!bad)
    if (any(bad)) {
        cat("NA or numerical zeros or ones encountered in fitted probabilities\n")
        cat(paste("dropping these", sum(bad), "cases, but proceed with caution\n"))
    }
    aic.factor <- (k1 - k2)/neff
    bic.factor <- (k1 - k2)/(2 * neff) * log(neff)
    v <- rep(NA, 3)
    arg1 <- matrix(m[!bad], nrow = neff, ncol = 3, byrow = FALSE)
    arg2 <- matrix(c(0, aic.factor, bic.factor), nrow = neff, 
        ncol = 3, byrow = TRUE)
    num <- arg1 - arg2
    s <- apply(num, 2, sd)
    numsum <- apply(num, 2, sum)
    v <- numsum/(s * sqrt(neff))
    names(v) <- c("Raw", "AIC-corrected", "BIC-corrected")
    pval <- rep(NA, 3)
    msg <- rep("", 3)
    for (j in 1:3) {
        if (v[j] > 0) {
            pval[j] <- 1 - pnorm(v[j])
            msg[j] <- "model1 > model2"
        }
        else {
            pval[j] <- pnorm(v[j])
            msg[j] <- "model2 > model1"
        }
    }
    out <- data.frame(v, msg,(pval))
    names(out) <- c("Vuong z-statistic", "H_A", "p-value")

    return(out)
}
###################

###########
### the D&L score test for overdispersion:
lambdahat <- yhat <- predict(PoissonGLM, type="response")
T_1 <- sum((y-lambdahat)^2 - y)/ sqrt(2*sum(lambdahat^2))
LRT_pval <- DLtest_pval <-  pnorm(T_1, lower.tail=FALSE)

###########
### the Vuong test for zero-inflation:
vuong_P_ZIP_pval <- vuong_NB_ZINB_pval <- 1

if(exists("zip_mod") & sum(y==0)>1 ){
	tryCatch({	
vv_P_ZIP<-(vuong_f(PoissonGLM, zip_mod))
vuong_P_ZIP_pval <-as.numeric(as.character(vv_P_ZIP[1,3]))
}, 
 error=function(e){})}

if(exists("zinb_mod")  & sum(y==0)>1 ){
		tryCatch({	
vv_NB_ZINB<-(vuong_f(NB_mod, zinb_mod))
vuong_NB_ZINB_pval <-as.numeric(as.character(vv_NB_ZINB[1,3]))
}, 
 error=function(e){})}
#######################################################

\end{verbatim}
\end{footnotesize}

\section{Appendix Figures}

\begin{figure}[p]
    \centering
\includegraphics[width=14cm]{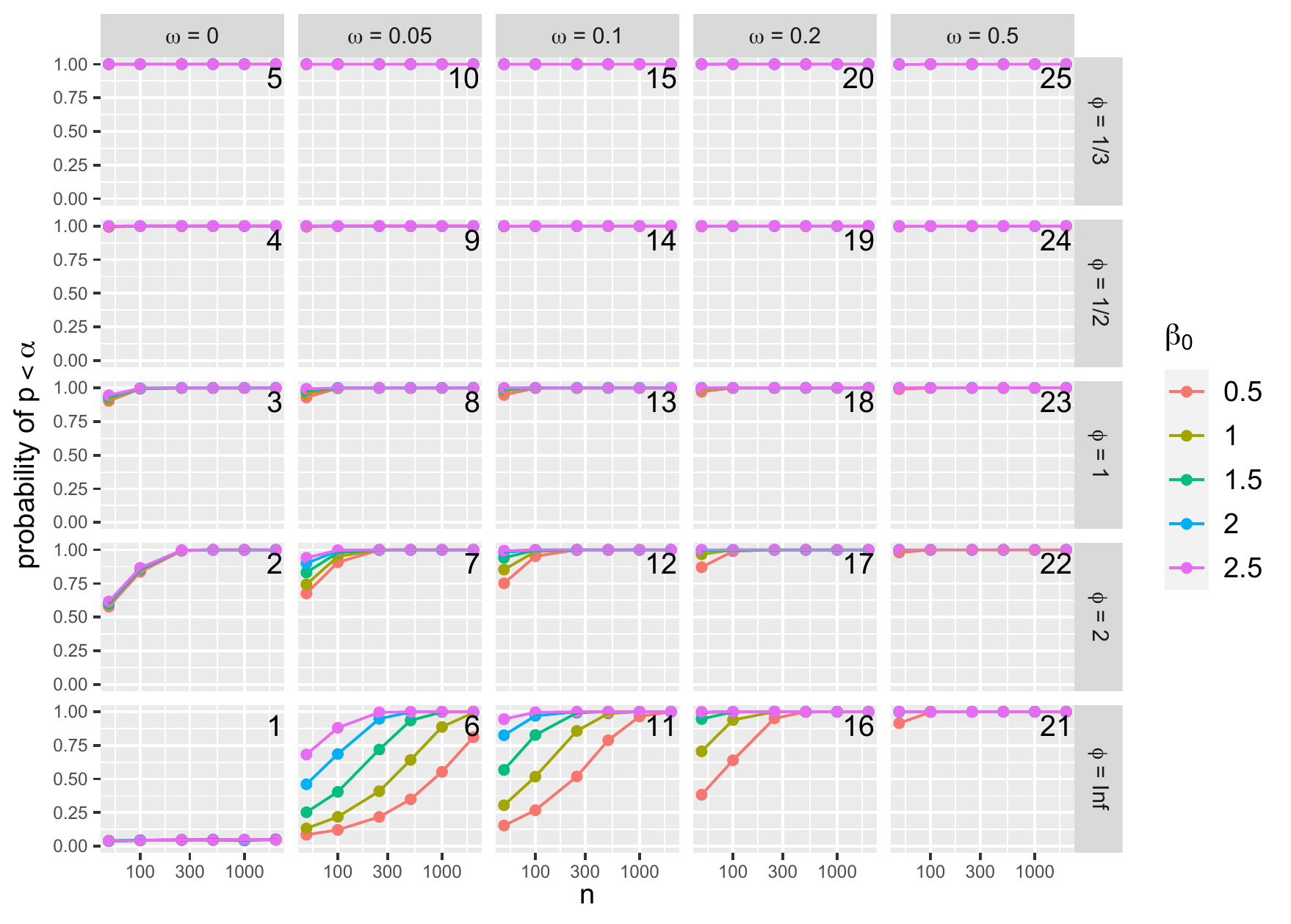}
    \caption{Probability that the D\&L test rejects the null hypothesis that there is no overdispersion.}
    \label{fig:overdispersion}
\end{figure}

\begin{figure}[p]
    \centering
\includegraphics[width=14cm]{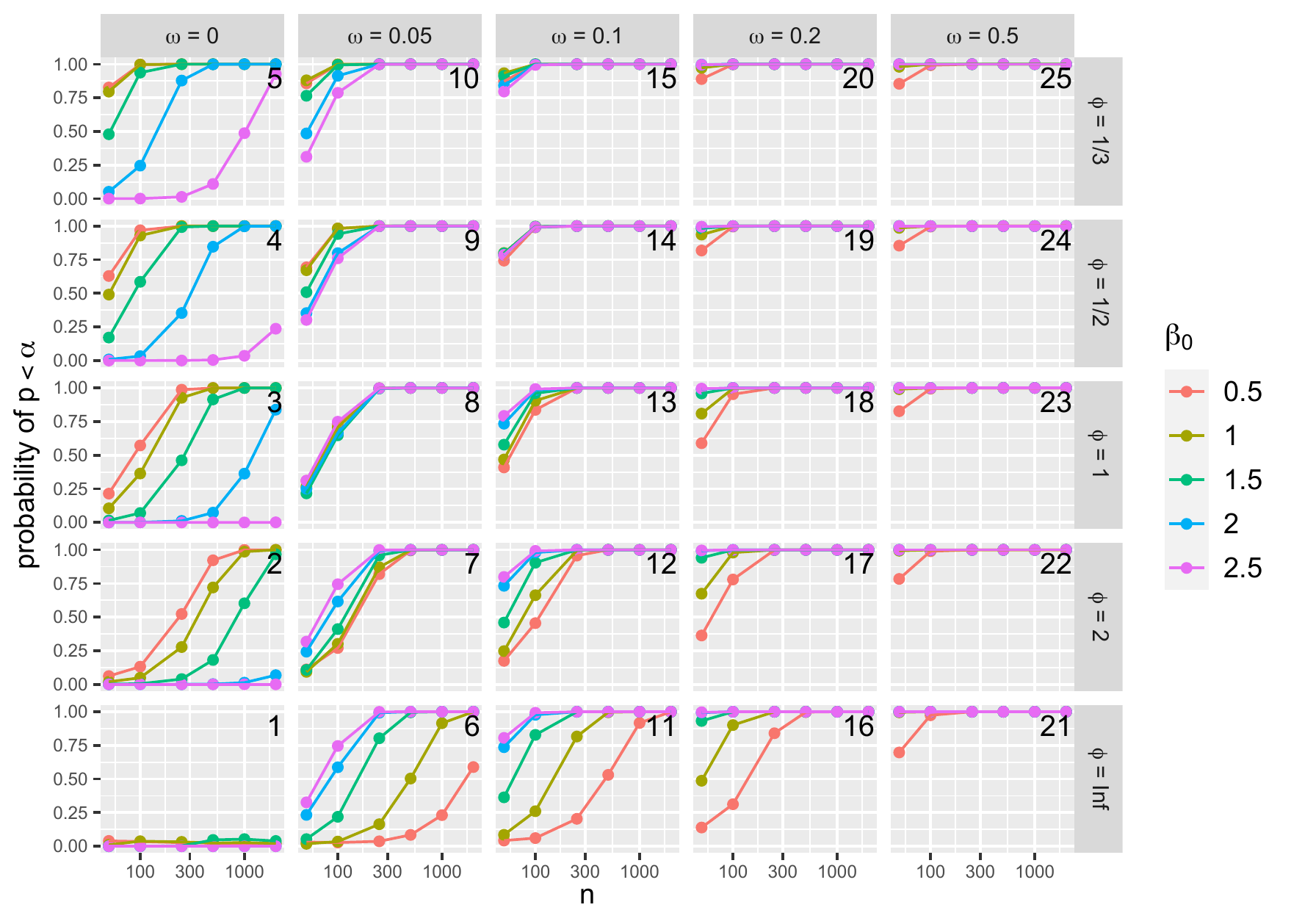}
    \caption{Probability that the Vuong test rejects the null hypothesis that there is no zero-inflation, comparing the Poisson model to the ZIP model.}
    \label{fig:vuongP}
\end{figure}

\begin{figure}[p]
    \centering
\includegraphics[width=14cm]{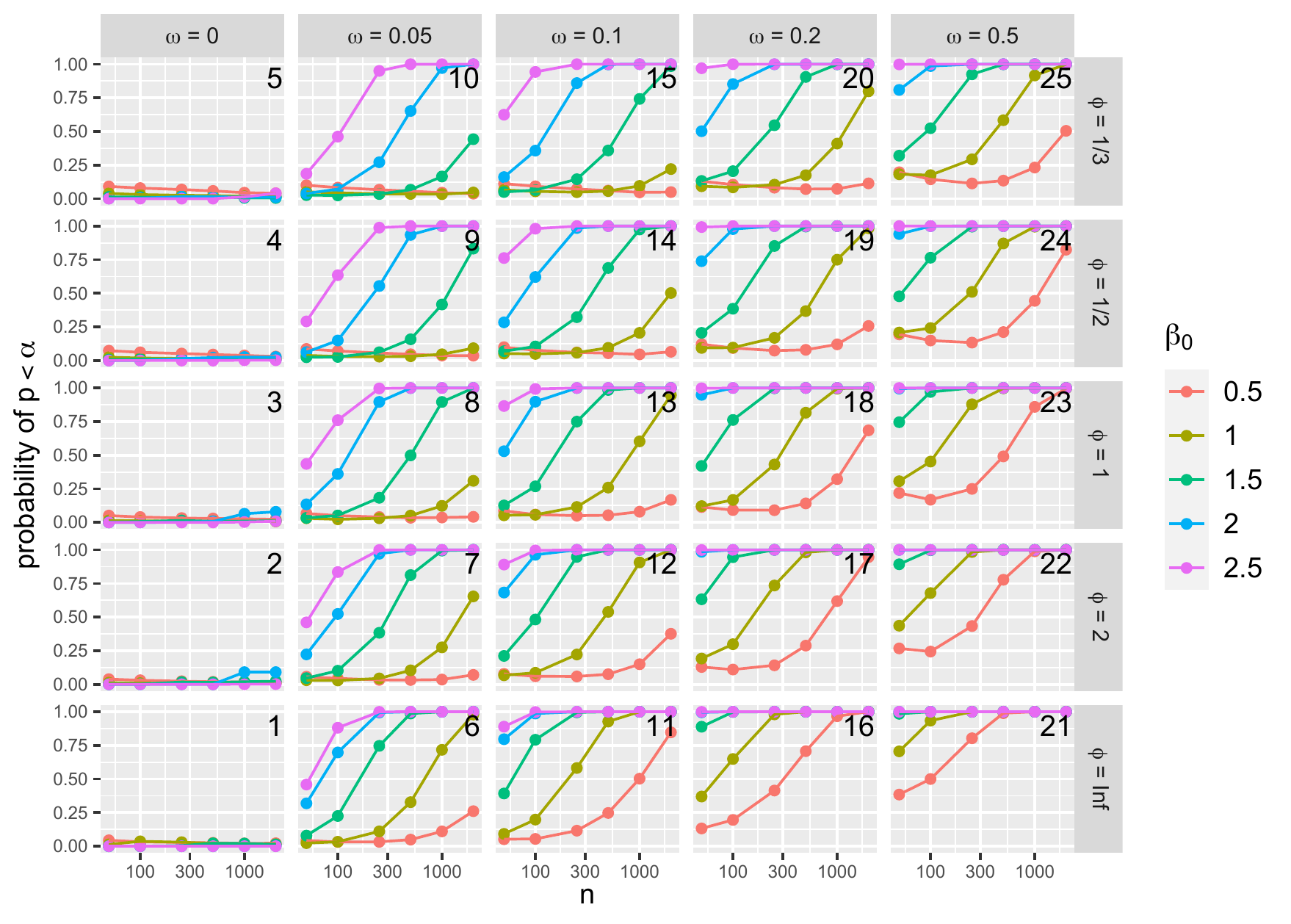}
    \caption{Probability that the Vuong test rejects the null hypothesis that there is no zero-inflation, comparing the NB model to the ZINB model.}
    \label{fig:vuongNB}
\end{figure}

\begin{figure}[p]
    \centering
\includegraphics[width=14cm]{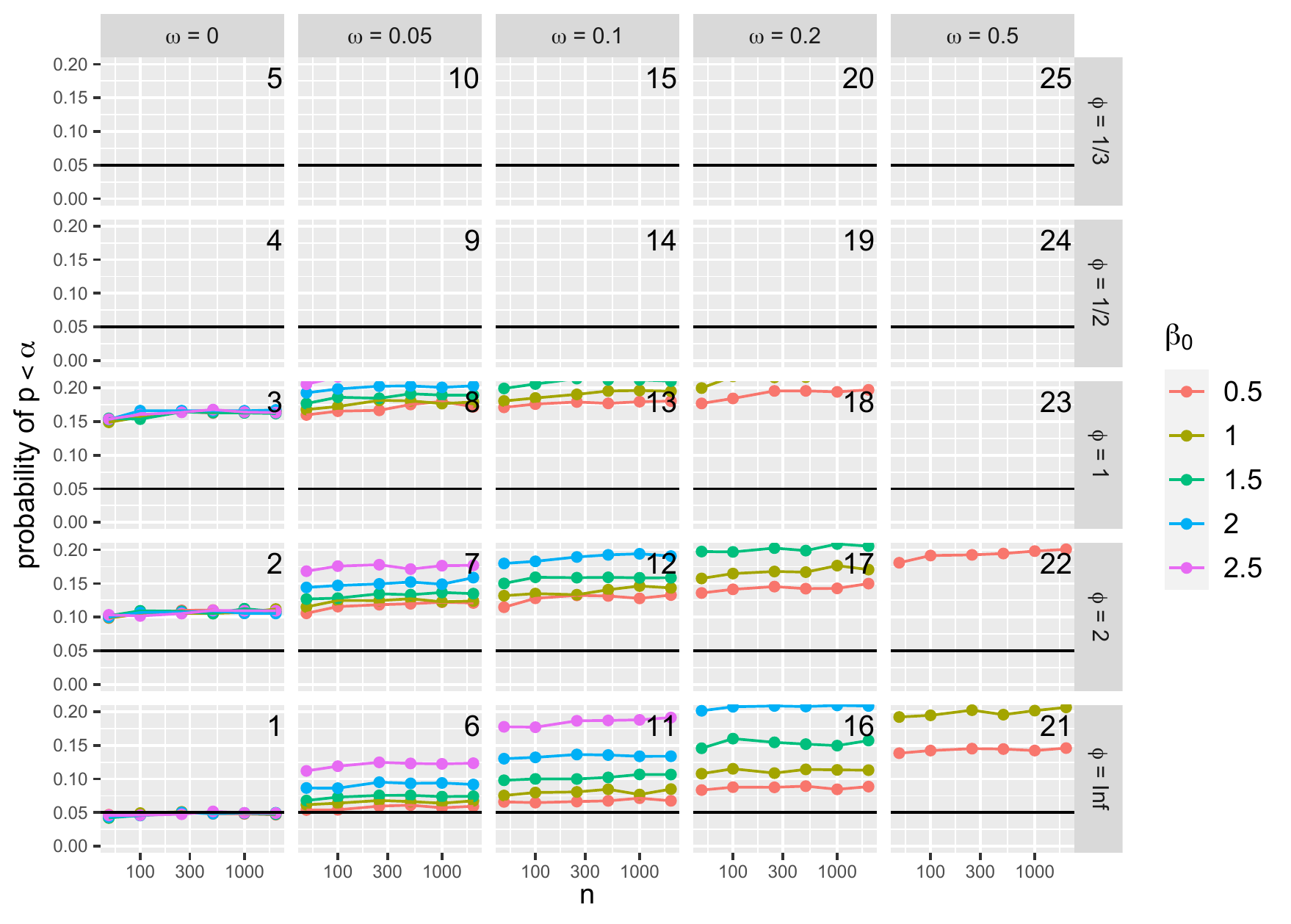}
    \caption{Probability that the Poisson model rejects the null hypothesis of $H_{0}: \beta_{X}=0$.}
    \label{fig:poisson}
\end{figure}

\begin{figure}[p]
    \centering
\includegraphics[width=14cm]{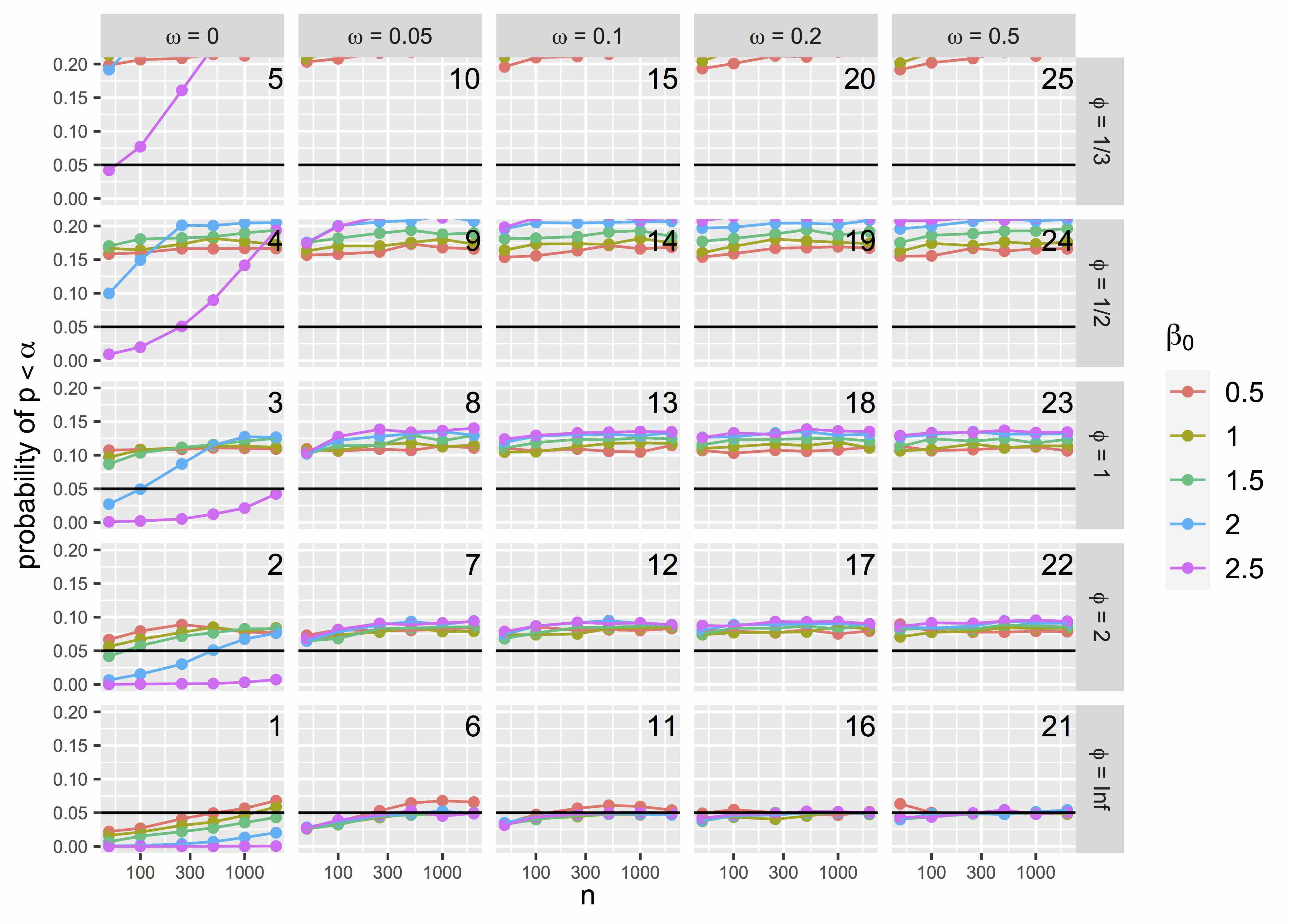}
    \caption{Probability that the ZIP model rejects the null hypothesis of $H_{0}: \beta_{X}=0$.}
    \label{fig:zip}
\end{figure}

\begin{figure}[p]
    \centering
\includegraphics[width=14cm]{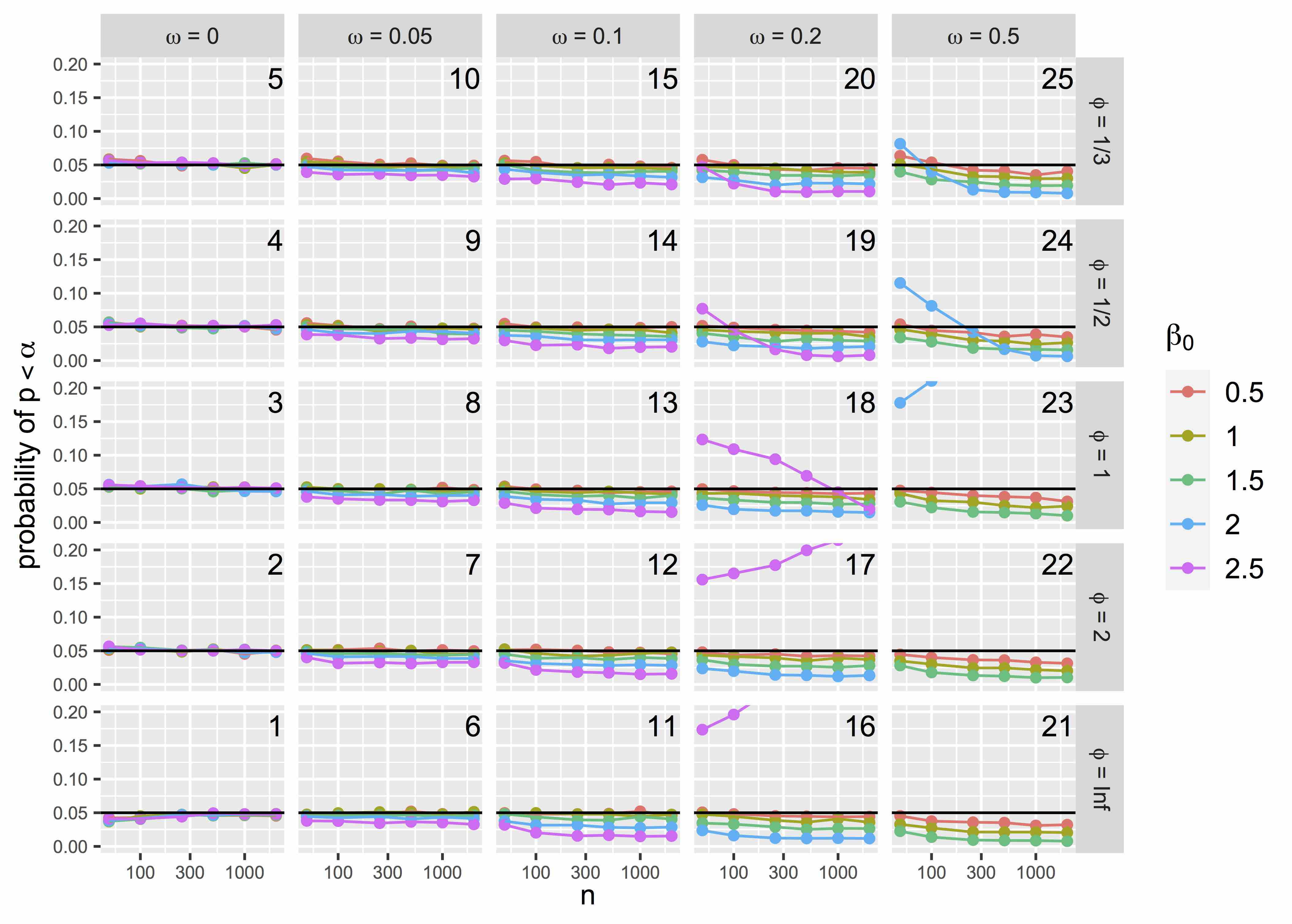}
    \caption{Probability that the NB model rejects the null hypothesis of $H_{0}: \beta_{X}=\gamma_{X}=0$.}
    \label{fig:nb}
\end{figure}

\begin{figure}[p]
    \centering
\includegraphics[width=14cm]{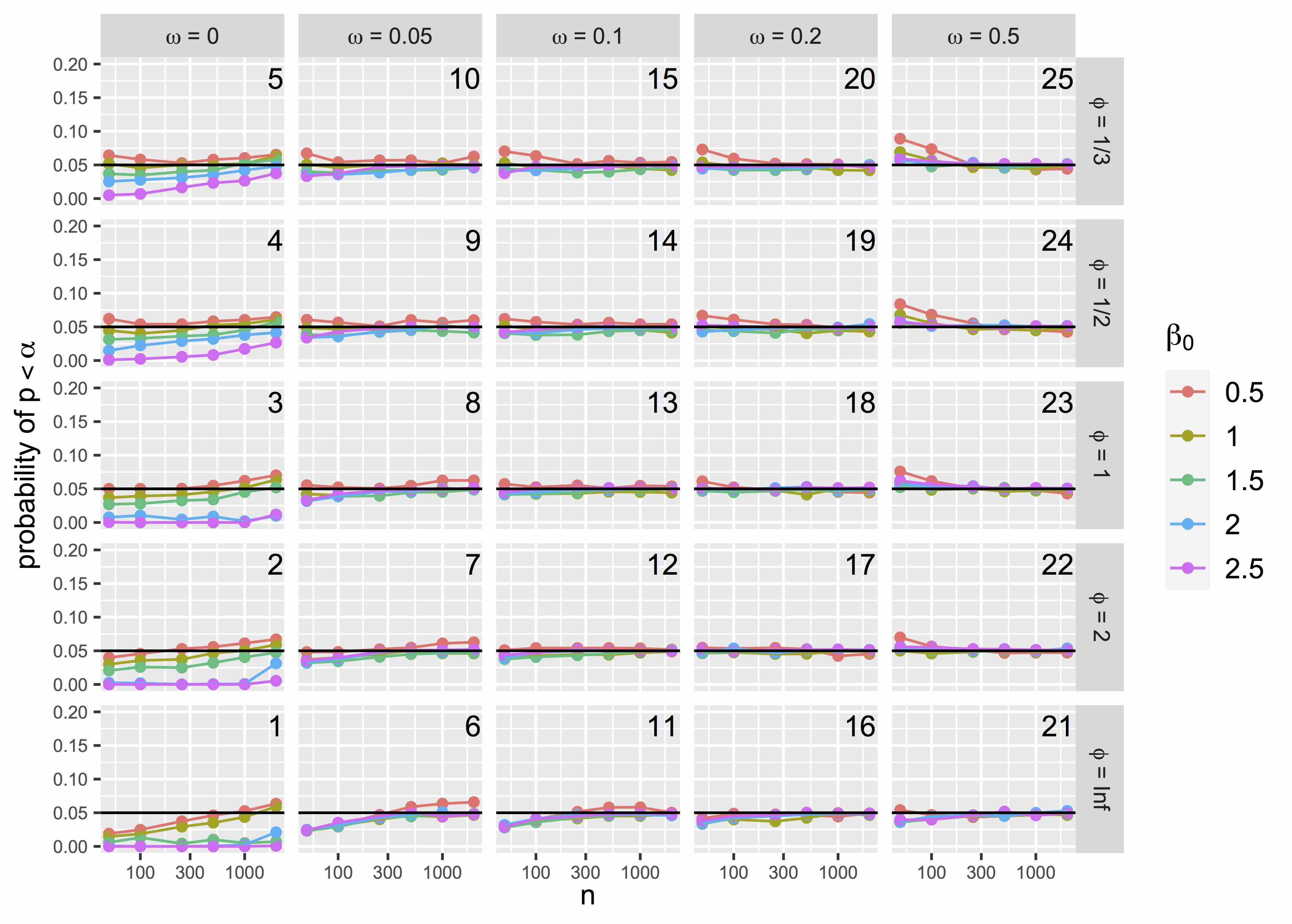}
    \caption{Probability that the ZINB model rejects the null hypothesis of $H_{0}: \beta_{X}=\gamma_{X}=0$.}
    \label{fig:zinb}
\end{figure}


\begin{figure}[p]
    \centering
\includegraphics[width=14cm]{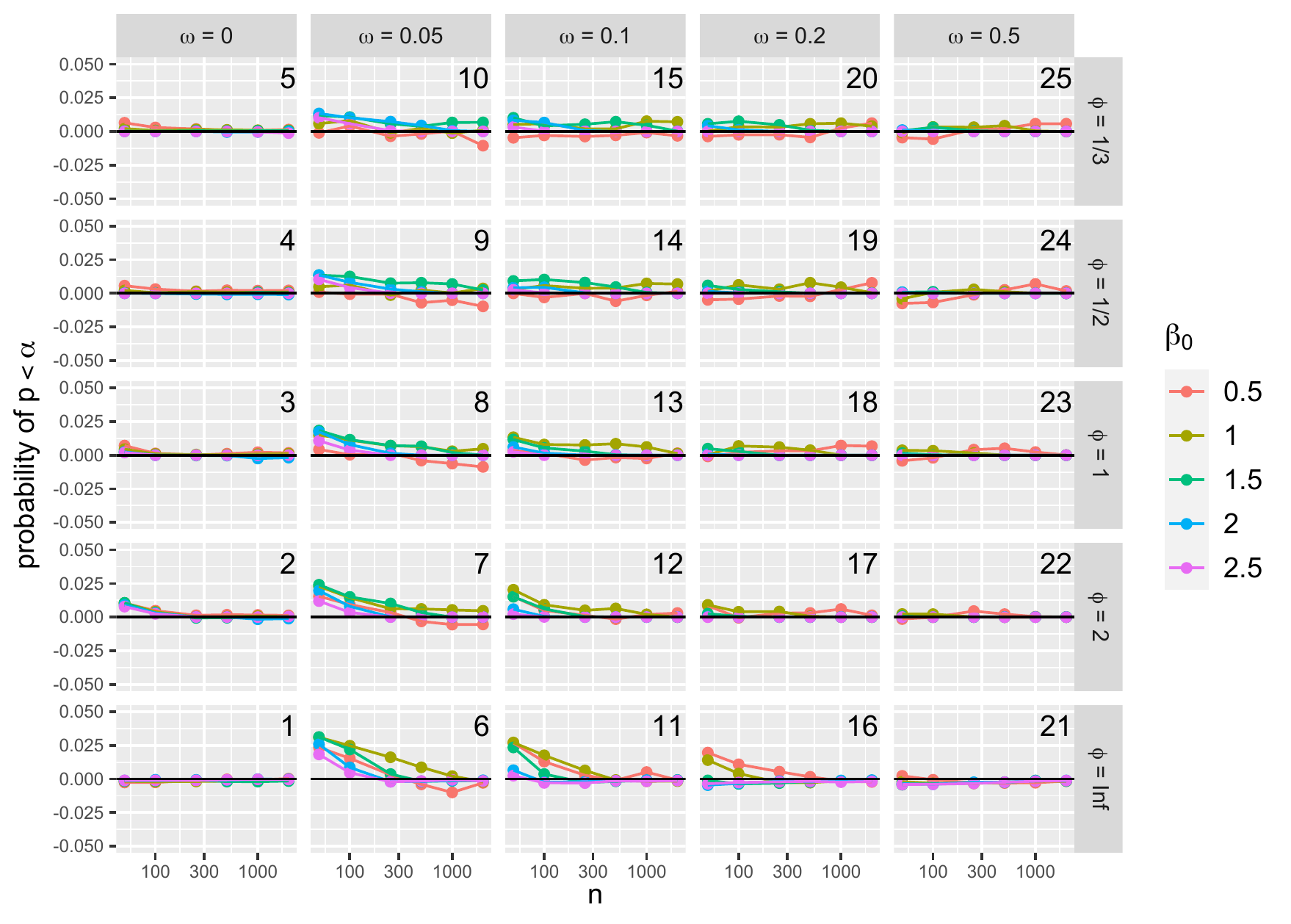}
    \caption{Difference between type 1 error under ``correct'' model (in Figure \ref{fig:correct}) and unconditional type 1 error (in Figure \ref{fig:type1}).}
    \label{fig:diff}
\end{figure}

\begin{figure}[p]
    \centering
\includegraphics[width=14cm]{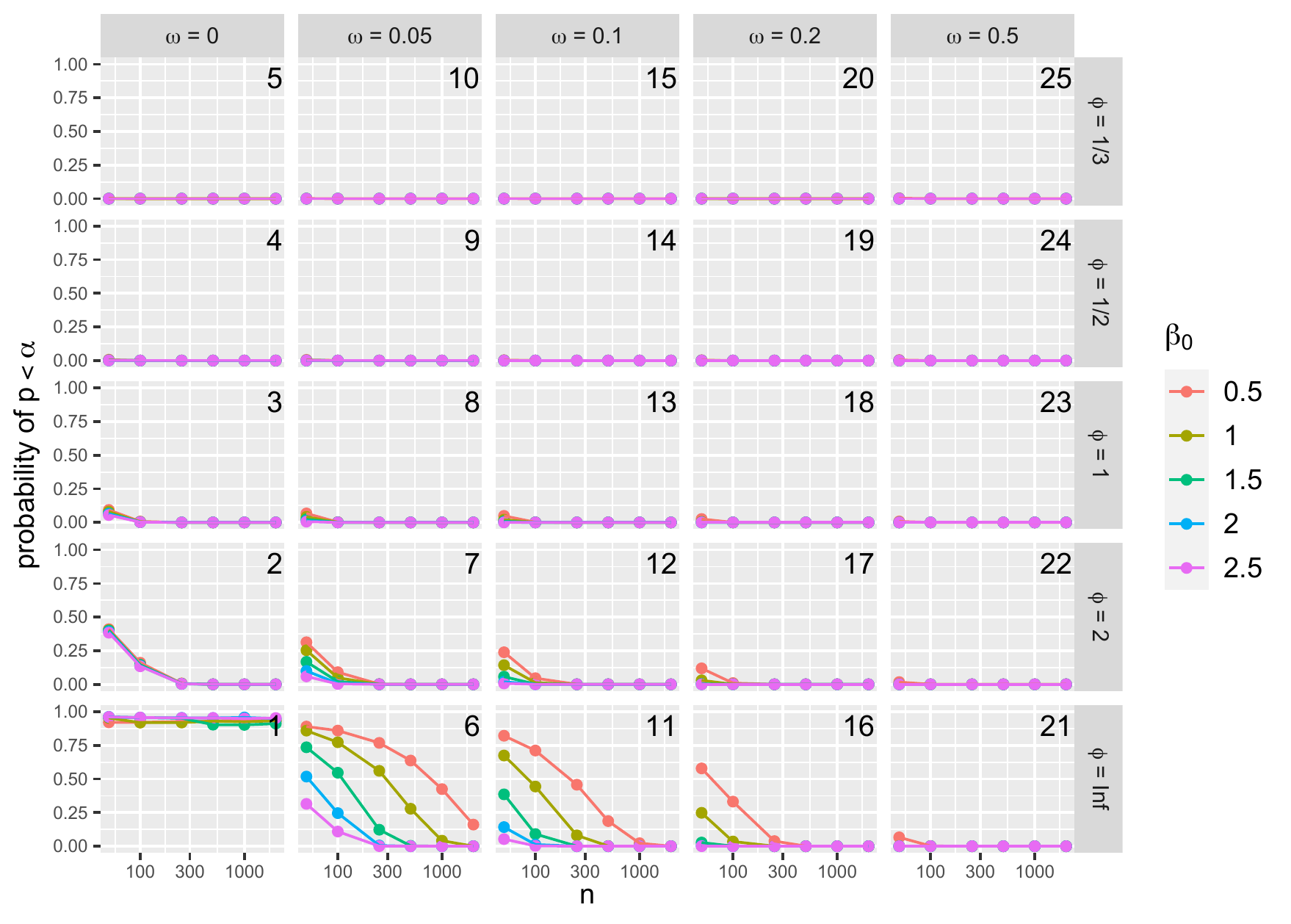}
    \caption{Proportion of datasets for which the preliminary testing scheme selects the Poisson model for analysis.}
    \label{fig:propPoisson}
\end{figure}

\begin{figure}[p]
    \centering
\includegraphics[width=14cm]{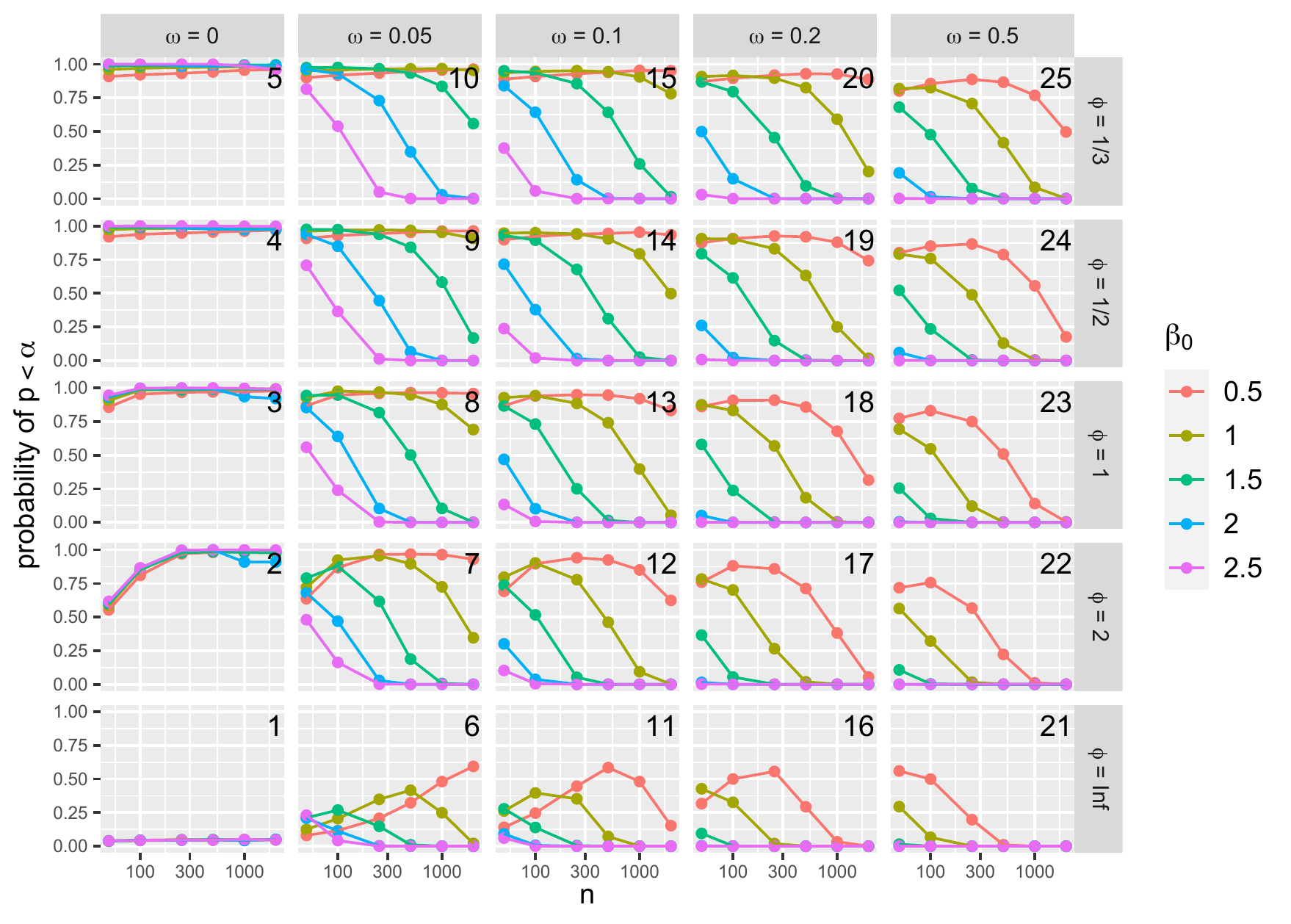}
    \caption{Proportion of datasets for which the preliminary testing scheme selects the NB model for analysis.}
    \label{fig:propNB}
\end{figure}

\begin{figure}[p]
    \centering
\includegraphics[width=14cm]{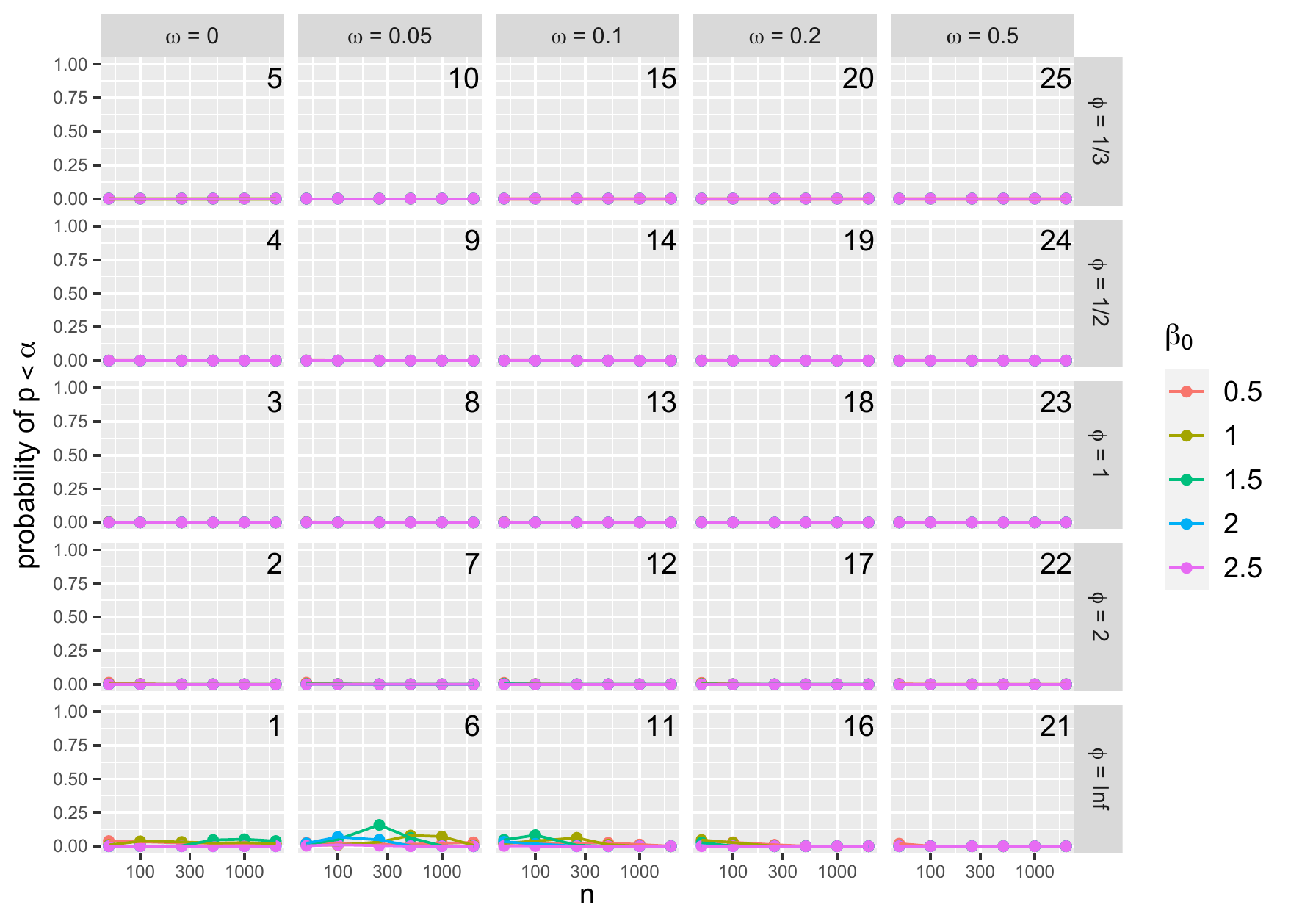}
    \caption{Proportion of datasets for which the preliminary testing scheme selects the ZIP model for analysis.}
    \label{fig:propZIP}
\end{figure}

\begin{figure}[p]
    \centering
\includegraphics[width=14cm]{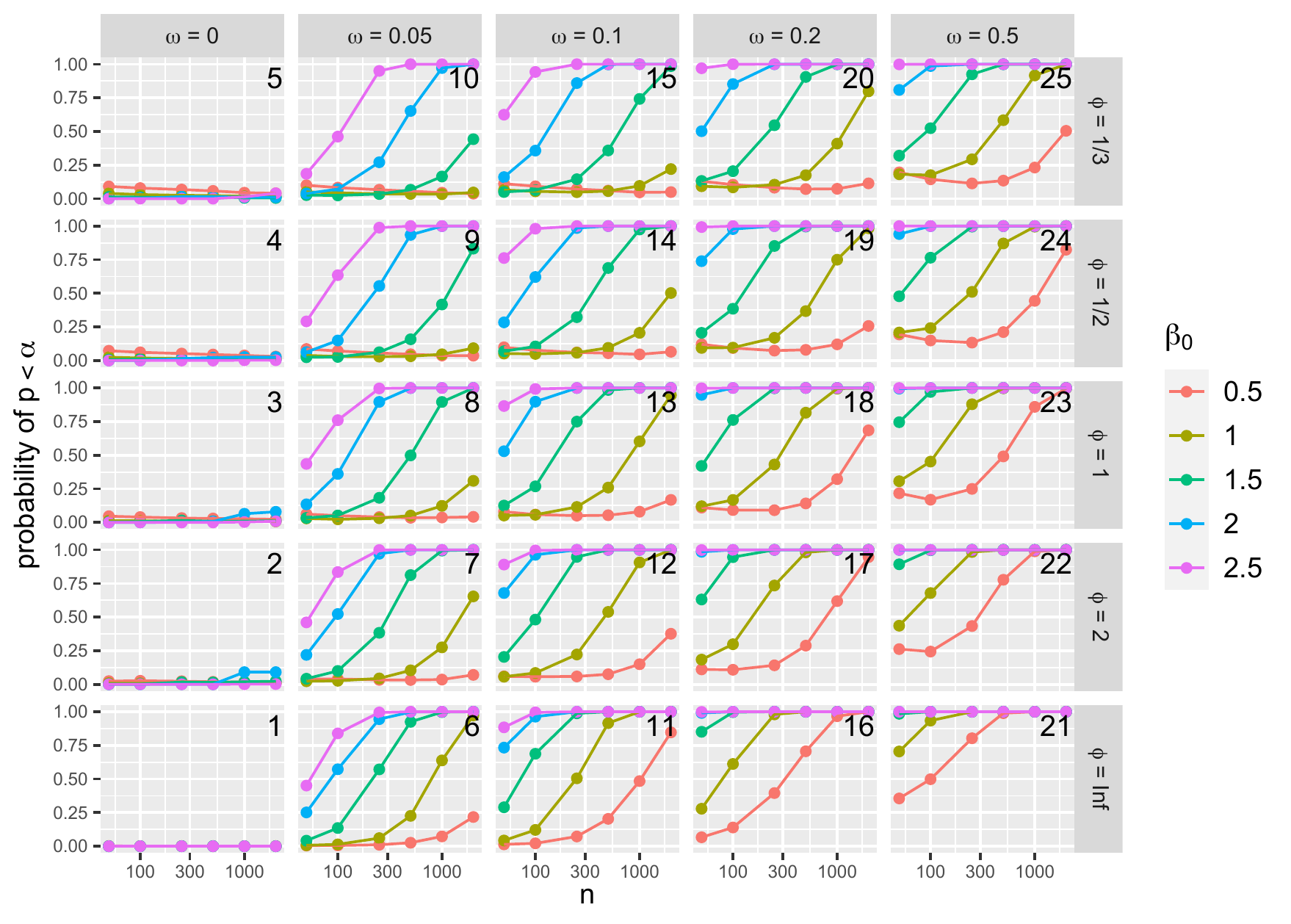}
    \caption{Proportion of datasets for which the preliminary testing scheme selects the ZINB model for analysis.}
    \label{fig:propZINB}
\end{figure}

\end{document}